\documentclass[%
 reprint,
 nofootinbib,
superscriptaddress,
 amsmath,amssymb,
 aps,
]{revtex4-2}
\usepackage[table]{xcolor}
\usepackage{graphicx}%
\usepackage{dcolumn}%
\usepackage{bm}%
\usepackage{braket}
\usepackage{physics}
\usepackage{tabularx}
\usepackage[inline]{asymptote}
\usepackage{siunitx}
\usepackage{tikz-cd}
\usetikzlibrary{backgrounds}
\usepackage{amsthm}

\usepackage[colorlinks=true,linkcolor=blue,allcolors=blue]{hyperref}
\usepackage[capitalize,nameinlink]{cleveref}

\begin{document}
\title{Low-Overhead Entangling Gates from Generalised Dehn Twists}%

\author{Ryan Tiew}
\affiliation{School of Mathematics, Fry Building, Woodland Road, Bristol BS8 1UG, UK}
\affiliation{Quantum Engineering Centre for Doctoral Training, University of Bristol, BS8 1FD, UK}
\author{Nikolas P.\ Breuckmann}
\affiliation{School of Mathematics, Fry Building, Woodland Road, Bristol BS8 1UG, UK}

\date{\today}%

\begin{abstract}
We generalise the implementation of logical quantum gates via Dehn twists from topological codes to the hypergraph and balanced products of cyclic codes. 
These generalised Dehn twists implement logical entangling gates with no additional qubit overhead and $\mathcal{O}(d)$ time overhead.
Due to having more logical degrees of freedom in the codes, there is a richer structure of attainable logical gates compared to those for topological codes.
To illustrate the scheme, we focus on families of hypergraph and balanced product codes that scale as $[[18q^2,8,2q]]_{q\in \mathbb{N}}$ and $[[18q,8,\leq 2q]]_{q\in \mathbb{N}}$ respectively.
For distance 6 to 12 hypergraph product codes, we find that the set of twists and fold-transversal gates generate the full logical Clifford group. For the balanced product code, we show that Dehn twists apply to codes in this family with odd $q$.
We also show that the $[[90,8,10]]$ bivariate bicycle code from \cite{bravyi2023highthreshold} is a member of the balanced product code family that saturates the distance bound.
We also find balanced product codes that saturate the bound up to $q\leq8$ through a numerical search. %

\end{abstract}

\maketitle

\section{Introduction}

In recent years, quantum low-density parity check (qLDPC) codes \cite{Breuckmann_2021} have seen significant theoretical and experimental breakthroughs. They are actively pursued in numerous physical architectures, such as in superconducting qubits \cite{bravyi2023highthreshold, acharya2024quantumerrorcorrectionsurface}, neutral atom arrays \cite{Bluvstein2024,xu2024fastparallelizablelogicalcomputation}, and ion-trap quantum computers \cite{ryananderson2022implementingfaulttolerantentanglinggates,PhysRevLett.133.180601}. 
However, not many schemes for implementing logical gates in qLDPC codes are known, and those that exist come with various drawbacks \cite{Jochym_O_Connor_2019, PhysRevX.11.011023,Cohen_2022,breuckmann2022foldtransversal,Quintavalle2023partitioningqubits, cross2024improvedqldpcsurgerylogical}. 
The scheme in \cite{Cohen_2022}, for example, suffers from large overhead due to needing ancillary qubits; and in \cite{breuckmann2022foldtransversal}, although gates have low overhead, not all Clifford gates can be implemented, as we are constrained by the available code symmetries. 
In general, it is unclear how to implement fault-tolerant logic in qLDPC codes, and therefore schemes for fault-tolerant logic and codes equipped with them are sought-after.

Dehn twists are a type of topological deformation that act on a surface. 
Roughly speaking, a twist is implemented by cutting the surface along a loop, rotating one side by $2\pi$, and then gluing the surface back together. 
Twisting along \textit{essential cycles}, which can be thought of as non-contractible loops, has the action of adding the cycle to others that cross it.
\begin{figure}[h!]
    \centering
    \includegraphics[width=\linewidth]{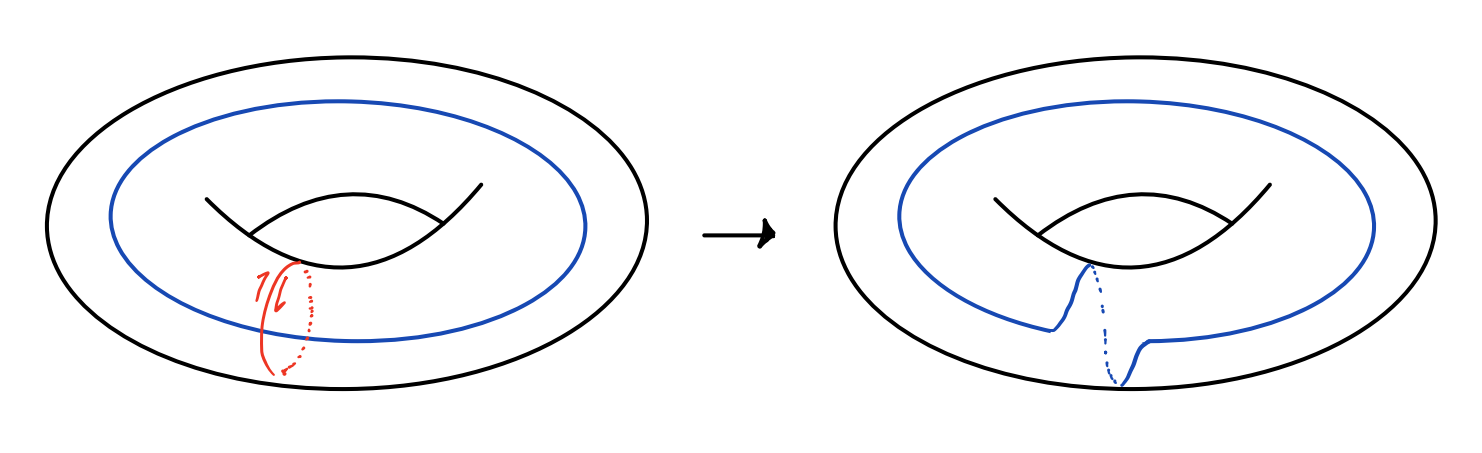}
    \caption{A Dehn twist on a torus. The torus has two essential cycles, one looping vertically (red) and one horizontally (blue). Here, a Dehn twist is illustrated by cutting the torus along the red loop, rotating it by $2\pi$, and gluing it back together. This adds the red loop to the blue one.}
    \label{fig:dehn twist picture for intro}
\end{figure}

Topological codes such as the toric code \cite{Kitaev_2003} cellulate a surface and take three consecutive terms in the associated chain complex as the quantum code. 
In these codes, essential cycles of the surface are logical operators, and Dehn twists can be used to implement logical operations via code deformations \cite{Koenig_2010,Breuckmann_2017, breuckmann2018phdthesishomologicalquantum,Lavasani_2019}. 
In the toric code, for example, Dehn twists act as $\operatorname{CNOT}$ gates between the logical qubit looping horizontally and the one looping vertically along the handle. 
The twists can be implemented with no additional qubit overhead, and $\mathcal{O}(d)$ time overhead, where $d$ is the code distance.

More generally, on a topological surface with $g$ handles, Dehn twists are orientation-preserving homeomorphisms that are known to generate the so-called mapping class group of the surface~\cite{Farb_Margalit_2017}. 
This group has a surjective representation on the symplectic group $\operatorname{Sp}(2g,\mathbb{F}_2)$ acting on the surface homology group -- the space of equivalent essential cycles -- which has dimension $2g$. 
Hence, on topological codes derived from these surfaces, Dehn twists provide a low-overhead method to implement logical quantum gates.

In this paper, we present two main pieces of work. Firstly, we propose a generalised Dehn twist scheme that can be applied beyond topological codes, to quantum codes obtained from the hypergraph \cite{Tillich_2014} and balanced products \cite{Breuckmann_2021_2} of cyclic codes. 
On these codes, Dehn twists can be implemented with low overhead, and error correction can be done in-between steps on the intermediate codes, which can have larger check weights and lower distances compared to the initial code. 
The generalised Dehn twists implement entangling operations on the logical space of the codes without additional ancilla qubits, and with $\mathcal{O}(d)$ time overhead. These act simultaneously on overlapping logical operators, taking them as control qubits, and mapping them to other logical operators. 
We also show that the number of possible Dehn twists is in a certain sense larger than that for topological codes, which can allow us to generate a larger set of logical operations.

Secondly, we construct code families from the hypergraph and balanced products of cyclic codes with a certain structure.
The code families have parameters $[[18q^2,8,2q]]_{q\in\mathbb{N}}$ and $[[18q,8,\leq2q]]_{q\in\mathbb{N}}$ respectively. %
They are constructed by taking a product of the simplest cyclic code after the repetition code, and in a sense are the simplest extensions of the toric code. 
The check weight in this construction is 6, for both $X$- and $Z$-checks.
The check weight can be increased by choosing different input cyclic codes; this leaves the parameters of the hypergraph product-based construction invariant, but can in some cases increase the distance for the balanced product. We also use these codes to show how, as there are more logical degrees of freedom, we have access to a larger set of logical operations from the generalised Dehn twists as compared to for topological codes.

For the hypergraph product, we investigate $d\in\{6,8,10,12\}$ codes with check weight 6, and find that Dehn twists combined with fold-transversal gates \cite{breuckmann2022foldtransversal} generate the full logical Clifford group. 
As fold-transversal gates also do not require any ancilla qubits, on these codes we can implement the full logical Clifford group with no additional qubit overhead.
We find for a choice of %
suitable cyclic codes that the intermediate code distances are not reduced while the maximum check weight increases to 9. Further simulations up to $q\leq 200$ show this upper bound on the intermediate check weight holds. 

For the balanced product, we present results of a numerical search for codes that saturate the $2q$ distance bound. We find that increasing the code check weight from~6 to~8 from using different cyclic codes
improves the maximum-achievable distance. In this case we find codes that saturate the upper distance bound $2q$ for all $1\leq q\leq 8$.
We provide the recipe to construct other balanced product cyclic codes with a fixed number of logical qubits by taking the balanced product of cyclic codes, and factoring out an appropriate subgroup. 
We implement the Dehn twists numerically for a $[[90,8,10]]$ code, and find that intermediate distances reduce to a minimum of~6, while intermediate check weights increase up to a maximum of~16. 
We also show that the $[[90,8,10]]$ bivariate bicycle code from~\cite{bravyi2023highthreshold} is from this family of balanced product cyclic codes that saturates the distance bound. 
This immediately extends the generalised Dehn twists to the code.
Furthermore, our analysis gives an analytic way to obtain a minimum-weight generating set, which was previously found in \cite{bravyi2023highthreshold} only through a numerical search.

The rest of the paper is structured as follows. In \cref{sec: preliminaries}, we give background information on CSS codes and how they are related to homology. In \cref{sec:theory}, we analyse the toric code and define its construction as a tensor product of two one-term chain complexes derived from cyclic group algebras. 
Using this language, we also describe how the Dehn twists implement logical $\operatorname{CNOT}$ gates on the toric code. 
In \cref{sec:general}, we show how Dehn twists can be generalised to the hypergraph product of other cyclic codes, and show that they implement logical mappings given a basis that we state from considering homology of cyclic codes. 
In \cref{sec:example}, we give a hypergraph product code family with parameters $[[18q^2,8,2q]]$, and a subset of possible Dehn twists associated with it. In \cref{sec: balanced product}, we construct the balanced product code family with parameters $[[18q,8,\leq2q]]$ and present results from a numerical search for codes that saturate the distance bound.
We discuss how Dehn twists apply to balanced product cyclic codes, and analyse a $[[90,8,10]]$ balanced product code to illustrate this. 
We also define the equivalence between our code and the bivariate bicycle code with the same parameters. 
Finally, in \cref{sec:conclusion}, we discuss future avenues to explore and give a summary of our work.
\section{Preliminaries} \label{sec: preliminaries}
We first give some basic definitions of CSS codes and their relation to the homology of chain complexes.

\subsection{CSS Codes}
A stabiliser quantum error-correcting code is one whose codewords are defined by the set of states
\begin{equation}
    \mathcal{C}=\{\ket{\psi}\mid P\ket{\psi}=\ket{\psi}\forall P\in \mathcal{S}\},
\end{equation}
where $\mathcal{S}$ is an abelian subgroup of the Pauli group $\operatorname{P}_n$ that does not contain $-\operatorname{I}$~\cite{GOTTESMAN2006196}. 
CSS codes are a type of stabiliser code where $\mathcal{S}$ can be generated by tensor products of Pauli operators acting as either $\operatorname{X}$ or $\operatorname{Z}$ only. 
They can be constructed from two classical binary linear codes, $C_X$ and $C_Z$, which detect phase-flip and bit-flip errors respectively. These codes can be defined by their parity check matrices $H_{\operatorname{X}}$ and $H_{\operatorname{Z}}$, where each row in the check matrix indicates the support of an $\operatorname{X}$ or $\operatorname{Z}$ stabiliser generator; these are also known as $\operatorname{X}$- and $\operatorname{Z}$-checks. 
If the row and column weights of the check matrices are bounded by a constant, then these are \textit{quantum low-density parity check} (qLDPC) codes. 
The codes satisfy the relation
\begin{equation} \label{eqn: Z dual in X}
        C_Z^\perp \subseteq C_X
\end{equation}
where $C_\alpha^\perp$ refers to the \textit{dual code} comprising all vectors $c^\perp$ that are orthogonal to vectors in $C_\alpha$. 
Equivalently, the parity check matrices of a CSS code must satisfy the property
\begin{equation} \label{eqn: CSS HxHz commute}
    H_{\operatorname{X}} H_{\operatorname{Z}}^T = 0.
\end{equation}
\subsection{(Co)Chain Complexes and (Co)Homology}
A based chain complex is defined by a sequence of free modules $\{A_i\}$ over a ring, together with linear maps $\partial_i : A_i \to \mathcal{A}_{i-1}$, that satisfy
\begin{equation}\label{eqn: boundary of boundary}
    \partial_{i-1} \circ \partial_i = 0 \quad \forall i.
\end{equation}
The maps $\partial_i$ are known as \textit{boundary operators}. %
Elements of $A_i$ are $i$-chains, while elements of the basis are known as $i$-cells. 
The $i^{th}$ homology group of the chain is defined as
\begin{equation}
    \mathcal{H}_i=\ker\partial_{i}/\operatorname{im}\partial_{i+1},
\end{equation}
which contains equivalence classes of \textit{essential cycles} in $A_i$: $i$-chains that do not have a boundary, but are themselves not the boundary of a chain in $A_{i+1}$. 
The two non-contractible loops of a torus are an example of essential cycles in $\mathcal{H}_1$.

In a similar fashion, we can also define the cochain complex by defining the coboundary operators $\delta_i:=\partial_{i+1}^T,\delta_i : A_{i} \to A_{i+1}$, and the cohomology group
\begin{equation}
    \mathcal{H}^i=\ker\delta_{i}/\operatorname{im}\delta_{i-1}.
\end{equation}
\subsection{Relation to CSS and Classical Codes}
Classical binary codes and quantum CSS codes have a natural relation to chain complexes when the set $\{A_i\}$ are defined over the binary field $\mathbb{F}_2$. 
A classical binary code can be interpreted as a 2-term chain complex 
\begin{center}
\begin{tikzcd}[cells={nodes={minimum height=2em}}]
A = ( \{0\} \arrow[r, "\partial_2"] & A_1\arrow[r, "\partial_1"] & A_0 \arrow[r, "\partial_0"] & \{0\} ),
\end{tikzcd}
\end{center}
where $\partial_1$ is the classical parity check matrix, and by definition $\operatorname{im}\partial_2=0$ and $\ker\partial_0=A_0$. We define a classical code by $\ker{\partial_1}$, which is the homology group $\mathcal{H}_1$ of the chain complex. Then there is a correspondence between bits and $1$-cells, and checks and $0$-cells.

Similarly, a quantum CSS code $\mathcal{C}$ defined by parity check matrices $H_{\operatorname{X}}$ and $H_{\operatorname{Z}}$ can be associated with a 3-term chain complex
\begin{center}
\begin{tikzcd}[cells={nodes={minimum height=2em}}]
B = (\{0\} \arrow[r, "\partial_3"]& B_2 \arrow[r, "\partial_2"] & B_1\arrow[r, "\partial_1"] & B_0 \arrow[r, "\partial_0"] & \{0\} ),
\end{tikzcd}
\end{center}
where $\operatorname{im}\partial_3=0,\ker\partial_0=B_0$. We equate $\partial_2$ with $H_Z^T$, and $\partial_1$ with $H_X$. The condition in \cref{eqn: CSS HxHz commute} is guaranteed by \cref{eqn: boundary of boundary}. $\operatorname{X}$-checks are elements in $B_0$, $\operatorname{Z}$-checks are elements in $B_2$, and qubits are elements in $B_1$. $\operatorname{X}$ and $\operatorname{Z}$ stabilisers are identified with coboundaries $\operatorname{im}\delta_0$ and boundaries $\operatorname{im}\partial_2$ respectively.
Then, considering the condition in \cref{eqn: Z dual in X}, we can define a correspondence between the homology group $\mathcal{H}_1$ and the set of logical $\operatorname{Z}$ operators,~$\bar{\operatorname{Z}}$. 
By considering the cochain, we can also define a similar correpondence between the set of logical $\operatorname{X}$ operators,~$\bar{\operatorname{X}}$ and the cohomology group $\mathcal{H}^1$.
Using chain complex terminology, the code distance $d_Z$ for $\operatorname{Z}$ is equal to the length of the \textit{systole}; and the code distance~$d_X$ for $\operatorname{X}$ is equal to the length of the \textit{cosystole}.\footnote{See \cite[Section~2.3]{breuckmann2018phdthesishomologicalquantum} or \cite[Section II.B]{Breuckmann_2021} for more details on the connection between homological algebra and CSS codes.} For the rest of the paper, we will abuse this correspondence when referring to quantum CSS codes, and switch freely between the language of coding theory and homological algebra.

Homological algebra lets us interpret a sequence of 3 terms $\{C_{i+1},C_i,C_{i-1}\}$ from an arbitrary chain as a quantum CSS code, or 2 terms as a classical code. The toric code, for example, can be thought of in this way as the cellulation of a torus by squares, with 2-cells as square faces, 1-cells as vertical and horizontal edges, and 0-cells as vertices. We see in following sections that the language also lets us \textit{construct} longer chains from the tensor product of shorter chains, which provides a way to construct quantum CSS codes from classical binary codes. 
\section{Toric Code}\label{sec:theory}
The toric code is a CSS code constructed from two classical repetition codes. It can be thought of as a tensor product of two one-term chain complexes, which are associated with repetition codes from two cyclic group algebras, $\mathbb{F}_2C_l$ and $\mathbb{F}_2C_m$. 
In order to set up notation, we use the familiar example of the toric code to write down Dehn twists in homological language and with cyclic polynomials.

\subsection{Cyclic Codes and Chain Complexes}
We define the cyclic groups
\begin{subequations} \label{eqn: cyclic groups}
\begin{align}
      C_l&=\langle y\mid y^l=e_l\rangle, \\
      C_m &= \langle x\mid x^m=e_m\rangle.
\end{align}
\end{subequations}
The group algebra $\mathbb{F}_2G$ for group $G$ is the set of all linear combinations of finitely-many elements $g\in G$ with binary coefficients. It is a vector space with $g$ as basis vectors. In general, a group algebra element can be represented by
\begin{equation}
    \sum_{g\in G} a_g g, a_g\in\mathbb{F}_2.
\end{equation}
Making the equivalence between polynomials~${\mathbb{F}_2[a]/\langle a^n+1\rangle}$ and the cyclic group algebra~$\mathbb{F}_2C_n$, we can define the check and generator polynomials for the repetition code in terms of a cyclic group algebra. These are given as
\begin{subequations}
\begin{align}
      p_r(a)&=e_{|a|}+a, \\
      g_r(a) &= \sum_{i=0}^{|a|-1} a^i,
\end{align}
\end{subequations}
where subscript $r$ denotes the repetition code, and $|a|$ is the order of the generator $a$ for $a\in \{x,y\}$.
For any cyclic group algebra $\mathbb{F}_2C_n$, we can always factorise the polynomial
\begin{equation}
    x^n+e_n = p_r(x)g_r(x).
\end{equation}

Repetition codes are the simplest example of a \textit{cyclic code}. A cyclic code is one where, if $c=(c_0,c_1,\cdots,c_n)$ is a codeword, then so is the cyclic shift $(c_n,c_0,\cdots,c_{n-1})$. Elements of a $\mathbb{F}_2C_n$ cyclic group algebra can be interpreted as codewords, and generalising from the repetition code, cyclic codes can be defined by check and generator polynomials $p(x)$ and $g(x)$ that satisfy 
\begin{equation} \label{eqn: vanishing add term 0}
    p(x)g(x) = 0
\end{equation}
from the factorisation of the polynomial $x^n+e_n$.

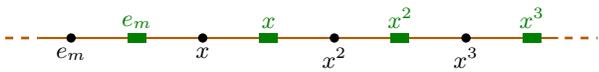
\begin{figure}
    \centering
    \begin{tikzpicture}[scale = 1.75, line width = 1pt]
    \colorlet{vertexcolor}{black} %
    \colorlet{linecolor}{orange!70!black}
    \colorlet{hcolor}{green!50!black} %
    \colorlet{vcolor}{blue!80!black}
    \def\dashstart{-0.5}
    \def\dashend{-0.25}
    \def\linestart{-0.25}
    \def\lineend{3.7}
    \def\scale{1}
    \def\squarewidth{0.035}

    \draw[thick, linecolor] (\linestart,0) -- (\lineend,0);
    \draw[dashed, linecolor] (\dashstart,0) -- (\dashend,0);
    \draw[dashed, linecolor] (3.75,0) -- (4,0);

    \foreach \x in {0,1,2,3} {
        \fill[black, anchor = north] (\x,0) circle (1pt);
    }
    
    \foreach \x in {0,1,2,3} {
        \fill[hcolor, anchor = south] (\x+0.5-2*\squarewidth,+\squarewidth) rectangle (\x+0.5+2*\squarewidth,-\squarewidth);
    }
    
    \def\hshift{0}
    
    \node[vertexcolor,anchor = north, scale = \scale] at (0+\hshift,0) {$e_m$};
    \node[vertexcolor,anchor = north, scale = \scale] at (1+\hshift,0) {$x$};
    \node[vertexcolor,anchor = north, scale = \scale] at (2+\hshift,0) {$x^2$};
    \node[vertexcolor,anchor = north, scale = \scale] at (3+\hshift,0) {$x^3$};

    \def\hshift{0.5}
    \node[anchor = south, scale=\scale, hcolor] at (0+\hshift,0) {$e_m$};
    \node[anchor = south, scale=\scale, hcolor] at (1+\hshift,0) {$x$};
    \node[anchor = south, scale=\scale, hcolor] at (2+\hshift,0) {$x^2$};
    \node[anchor = south, scale=\scale, hcolor] at (3+\hshift,0) {$x^3$};

\end{tikzpicture}
    \caption{Chain complex associated with the classical repetition code from the regular representation of the cyclic group algebra $\mathbb{F}_2C_m$. Checks are drawn as black vertices, while bits (edges in the chain complex) are drawn as green horizontal rectangles. Both are labelled by cyclic group algebra elements. There are $m$ vertices and edges as the regular representation of the group algebra elements are $m\times m$ matrices. The check polynomial $p_r(x) = e_m+x$ connects every edge labelled by $h\in C_m$ to vertices $p_r(x)h = h+hx$; these connections are shown as orange lines. Due to the cyclic group algebra, there are periodic boundary conditions, and so this is a circle. 
    The codeword is given by the set of green vertices $g_r(x)$, which is every edge in the circle.} 
    \label{fig:classical repn code}
\end{figure}

A cyclic code can be thought of in terms of a 2-term chain complex from cyclic group algebras
\begin{center}
\begin{tikzcd}[cells={nodes={minimum height=2em}}]
\mathbb{F}_2C_m\arrow[r, "\partial_A"] & \mathbb{F}_2C_m
\end{tikzcd}
\end{center}
where $\partial_A$ is a 1-dimensional matrix over the group algebra.
To obtain a code over $\mathbb{F}_2$, we switch to the \textit{regular representation} of the group algebra, where every group element is mapped to a $m\times m$ matrix over $\mathbb{F}_2$ representing its permutation action on other group elements. 
Taking the regular representation for $\partial_A=[p_r(a)]$ returns a square parity check matrix corresponding to that of the classical repetition code.\footnote{The canonical form of these parity check matrices is not square. Linearly-dependent rows are removed. 
It doesn't matter for classical codes, but for quantum codes this changes the dimension of the $0^{th}$ homology group, in turn affecting the number of logical qubits in the code. 
For example, the product of 2 circles (repetition codes) give a torus: the toric code. 
The product of a \textit{line}, removing the row in the repetition code parity check matrix that is not linearly-independent, and its dual, gives a patch: the surface code.} 
Associating the checks and bits of the binary code with 0-cells and 1-cells in the chain complex, checks can be thought of as vertices, while bits can be thought of as edges. This representation is shown in \cref{fig:classical repn code}.

For the toric code, we consider the \textit{double complex} from taking a tensor product of two such chains. A double complex $E$ constructed from chains $\{A_i\}$ and $\{B_j\}$ and their associated boundary operators has its terms indexed as $E_{i,j}$. The double complex for the toric code is given in \cref{fig:double_complex},
\begin{figure}
    \centering
\tikzcd[cells={nodes={minimum height=2em}}]
\mathbb{F}_2C_l\otimes\mathbb{F}_2C_m \arrow[d,"\partial_v"] \arrow[r, "\partial_h"] & \mathbb{F}_2C_l\otimes\mathbb{F}_2C_m \arrow[d,"\partial_v"]\\
\mathbb{F}_2C_l\otimes\mathbb{F}_2C_m \arrow[r, "\partial_h"] & \mathbb{F}_2C_l\otimes\mathbb{F}_2C_m
\endtikzcd,
    \caption{Double complex from the tensor product of two 2-term chains, each constructed from cyclic group algebras. We choose $\partial_v,\partial_h$ to represent vertical and horizontal boundary operators respectively. For a double complex $E$, each term can be indexed by $E_{i,j}$ representing the index of the original chain complexes the double complex was constructed from.
    }
    \label{fig:double_complex}
\end{figure}
where we can take
\begin{subequations}
    \begin{align}
        \partial_v &= [p_r(y)\otimes e_m],\\
        \partial_h &= [e_l\otimes p_r(x)].
    \end{align}
\end{subequations}
We adopt the notation that $l$ corresponds to the torus height and $m$ corresponds to the torus width. This associates $\partial_v$ with the vertical boundary operator, and $\partial_h$ with the horizontal boundary operator in the double complex. We also adopt the notation that 
\begin{subequations}
    \begin{align}
        \partial_v:E_{i,j}&\to E_{i-1,j},\\
        \partial_h:E_{i,j}&\to E_{i,j-1}.
    \end{align}
\end{subequations}
To construct a CSS code, we define a \textit{total complex}\footnote{The hypergraph product construction \cite{Tillich_2014} is the CSS code associated with such a total complex.} by collapsing the double complex along the diagonal into a 3-term chain complex. Terms in the total complex are defined as
\begin{equation}
    \operatorname{Tot}(E)_n = \bigoplus_{i+j=n}A_i\otimes B_j.
\end{equation}
By considering the total complex constructed from the double complex defined in \cref{fig:double_complex}, we can write down the boundary operators of the toric code as
\begin{subequations}\label{eqn:boundary}
\renewcommand\arraystretch{1.2}
\begin{align}
      \partial_1 &= \left[p_r(y)\otimes e_m\mid e_l\otimes p_r(x)\right],\\
      \partial_2 &=\left[ 
        \begin{array}{c}
            e_l\otimes p_r(x)\\\hline
             p_r(y)\otimes e_m
        \end{array}
        \right]
\end{align}
\end{subequations}
Again we switch to the regular representation to obtain the standard binary $\operatorname{X}$- and (transposed) $\operatorname{Z}$-check matrices associated with the toric code. For the rest of the paper, we leave implicit that group algebra elements in the boundary and coboundary operators are written in their regular representation to obtain parity check matrices over $\mathbb{F}_2$. Note that, in the regular representation, the matrices representing the group algebra elements have a single non-zero entry in each row and column describing its action on other group algebra elements. Therefore, the check weight of the quantum code is the number of terms of the polynomials in the boundary and coboundary operators. In the toric code,  this returns a check weight of 4. We make a note here that by using higher-weight representative polynomials for the repetition code, we can obtain a quantum code with a larger check weight but the same $[[n,k,d]]$ as the toric code.

\begin{figure}[h!]
\centering
\begin{tikzpicture}[scale=2.4, line width = 1pt]
    \colorlet{vertexcolor}{black} %
    \colorlet{linecolor}{orange!70!black}
    \colorlet{hcolor}{green!50!black} %
    \colorlet{vcolor}{blue!80!black}
    \colorlet{xvlog}{yellow!80!black}
    \colorlet{zvlog}{teal!80!black}
    \def\highlightwidth{2}
    \def\highlightopacity{0.5}
    \def\dashstart{-0.5}
    \def\dashend{-0.25}
    \def\linestart{-0.25}
    \def\lineend{2.7}
    \def\scale{1}
    \def\squarewidth{0.035}
    \def\trianglewidth{0.1}
    \def\root{1.73205080757}
    \foreach \x in {0,1,2} {
        \draw[thick, linecolor] (\x,\linestart) -- (\x,\lineend);
        \draw[dashed, linecolor] (\x,\dashstart) -- (\x,\dashend);
        \draw[dashed, linecolor] (\x,2.75) -- (\x,3);
    }
    
    \foreach \y in {0,1,2} {
        \draw[thick, linecolor] (\linestart,\y) -- (\lineend,\y);
        \draw[dashed, linecolor] (\dashstart,\y) -- (\dashend,\y);
        \draw[dashed, linecolor] (2.75,\y) -- (3,\y);
    }
    
    \foreach \x in {0,1,2} {
        \foreach \y in {0,1,2} {
            \fill[black] (\x,\y) circle (1pt);
        }
    }
    
    \foreach \x in {0,1,2} {
        \foreach \y in {0,1,2} {
            \fill[hcolor] (\x+0.5-2*\squarewidth,\y+\squarewidth) rectangle (\x+0.5+2*\squarewidth,\y-\squarewidth);
        }
    }

    \foreach \x in {0,1,2} {
        \foreach \y in {0,1,2} {
            \fill[vcolor] (\x-\squarewidth,\y+0.5+2*\squarewidth) rectangle (\x+\squarewidth,\y+0.5-2*\squarewidth);
        }
    }

    \def\hshift{0}
    \def\vshift{-0}
    
    \node[vertexcolor,anchor = north east, scale = \scale] at (0+\hshift,2+\vshift) {$(y^2,e_m)$};
    \node[vertexcolor,anchor = north east, scale = \scale] at (1+\hshift,2+\vshift) {$(y^2,x)$};
    \node[vertexcolor,anchor = north east, scale = \scale] at (2+\hshift,2+\vshift) {$(y^2,x^2)$};
    
    \node[vertexcolor,anchor = north east, scale = \scale] at (0+\hshift,1+\vshift) {$(y,e_m)$};
    \node[vertexcolor,anchor = north east, scale = \scale] at (1+\hshift,1+\vshift) {$(y,x)$};
    \node[vertexcolor,anchor = north east, scale = \scale] at (2+\hshift,1+\vshift) {$(y,x^2)$};
    
    \node[vertexcolor,anchor = north east, scale = \scale] at (0+\hshift,0+\vshift) {$(e_l,e_m)$};
    \node[vertexcolor,anchor = north east, scale = \scale] at (1+\hshift,0+\vshift) {$(e_l,x)$};
    \node[vertexcolor,anchor = north east, scale = \scale] at (2+\hshift,0+\vshift) {$(e_l,x^2)$};
    
    \def\hshift{0.5}
    \def\vshift{0}
    
    \node[anchor = south, scale=\scale, hcolor] at (0+\hshift,2+\vshift) {$(y^2,e_m)$};
    \node[anchor = south, scale=\scale, hcolor] at (1+\hshift,2+\vshift) {$(y^2,x)$};
    \node[anchor = south, scale=\scale, hcolor] at (2+\hshift,2+\vshift) {$(y^2,x^2)$};
    
    \node[anchor = south, scale=\scale, hcolor] at (0+\hshift,1+\vshift) {$(y,e_m)$};
    \node[anchor = south, scale=\scale, hcolor] at (1+\hshift,1+\vshift) {$(y,x)$};
    \node[anchor = south, scale=\scale, hcolor] at (2+\hshift,1+\vshift) {$(y,x^2)$};
    
    \node[anchor = south, scale=\scale, hcolor] at (0+\hshift,0+\vshift) {$(e_l,e_m)$};
    \node[anchor = south, scale=\scale, hcolor] at (1+\hshift,0+\vshift) {$(e_l,x)$};
    \node[anchor = south, scale=\scale, hcolor] at (2+\hshift,0+\vshift) {$(e_l,x^2)$};
    \def\hshift{0}
    \def\vshift{0.5}
    \node[anchor = west, scale=\scale, vcolor] at (0+\hshift,2+\vshift) {$(y^2,e_m)$};
    \node[anchor = west, scale=\scale, vcolor] at (1+\hshift,2+\vshift) {$(y^2,x)$};
    \node[anchor = west, scale=\scale, vcolor] at (2+\hshift,2+\vshift) {$(y^2,x^2)$};
    
    \node[anchor = west, scale=\scale, vcolor] at (0+\hshift,1+\vshift) {$(y,e_m)$};
    \node[anchor = west, scale=\scale, vcolor] at (1+\hshift,1+\vshift) {$(y,x)$};
    \node[anchor = west, scale=\scale, vcolor] at (2+\hshift,1+\vshift) {$(y,x^2)$};
    
    \node[anchor = west, scale=\scale, vcolor] at (0+\hshift,0+\vshift) {$(e_l,e_m)$};
    \node[anchor = west, scale=\scale, vcolor] at (1+\hshift,0+\vshift) {$(e_l,x)$};
    \node[anchor = west, scale=\scale, vcolor] at (2+\hshift,0+\vshift) {$(e_l,x^2)$};
\end{tikzpicture}
    \caption{Layout of the toric code, with $\operatorname{X}$-checks and qubits labelled by a tuple of group algebra elements. $\operatorname{X}$-checks are represented as black circles, while qubits are represented as vertical blue rectangles (vertical edges) and horizontal green rectangles (horizontal edges). Connections are given in orange, vertex labels are colored accordingly, and $\operatorname{Z}$-checks are omitted for clarity. The toric code check polynomials $(e_l+y)$ and $(e_m+x)$ connect the vertical and horizontal edges labelled $(y^i,x^k)$ to the $\operatorname{X}$-checks $\{(y^i,x^k),(y^{i+1},x^k)\}$ and $\{(y^i,x^k),(y^{i},x^{k+1})\}$ respectively. The two-dimensional layout can be generalised to the hypergraph product of cyclic codes, albeit with different connections between qubits and $\operatorname{X}$-checks depending on the check polynomials.}
    \label{fig:toric layout}
\end{figure}
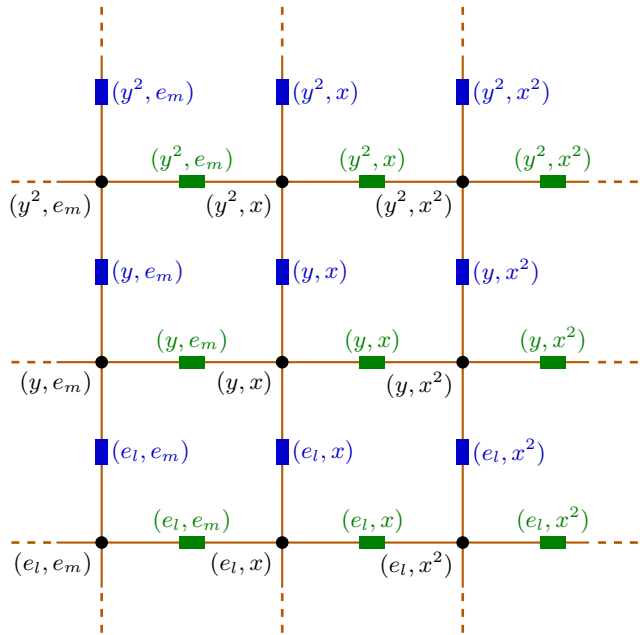

The connection between the toric code and classical cyclic codes is highlighted by working with polynomials. It is also apparent by considering the double complex as well as the space on which these boundary (coboundary) operators act that we can also label the $\operatorname{Z}$- ($\operatorname{X}$-) checks of the code, as well as the qubits, with group algebra elements. This is shown visually in \cref{fig:classical repn code} for the classical repetition code and \cref{fig:toric layout} for the toric code.

In the total complex, $\operatorname{Z}$- ($\operatorname{X}$-) checks correspond to $lm$ 2-cells and 0-cells respectively, indexed by elements of the group algebra.
There are $2lm$ qubits corresponding to 1-cells of the total complex:
$lm$ qubits correspond to cells of $E_{1,0}$ of the original double complex, interpreted as vertical edges, while the other $lm$ qubits are cells of $E_{0,1}$, interpreted as horizontal edges.\footnote{The standard definition of the hypergraph product as in \cite{Tillich_2014} defines qubits as cells of $E_{1,1}$ and $E_{0,0}$. This is simply a different way to define the total complex. We adopt the notation in this paper to make consistent the fact that qubits are defined by terms in the double complex with the same grading in the total complex.}
We work with notation that for row vectors written as $[a|b]$, $a,b\in \mathbb{F}_2C_l\otimes\mathbb{F}_2C_m$, $a$ corresponds to vertical edges, while $b$ corresponds to horizontal edges for the rest of the paper. This is simply a choice in how to stack $\partial_h,\partial_v$ to construct the total complex boundary operators given in \cref{eqn:boundary}.

\begin{figure}
    \centering
    \begin{tikzpicture}[scale = 1.75, line width = 1pt]
    
    \colorlet{vertexcolor}{black} %
    \colorlet{linecolor}{orange!70!black}
    \colorlet{hcolor}{green!50!black} %
    \colorlet{vcolor}{blue!80!black}
    \def\dashstart{-0.45}
    \def\dashend{-0.25}
    \def\linestart{-0.25}
    \def\lineend{3.7}
    \def\scale{1}
    \def\squarewidth{0.035}
    
    \clip (\dashstart,-0.25) rectangle (\lineend-\dashend,0.3);
    \draw[thick, linecolor] (\linestart,0) -- (\lineend,0);
    \draw[dashed, linecolor] (\dashstart,0) -- (\dashend,0);
    \draw[dashed, linecolor] (3.75,0) -- (4,0);
    \draw [linecolor] plot [smooth, tension = 1] coordinates {(0.5,0) (1.25,0.15) (2,0)};
    \draw [linecolor] plot [smooth, tension = 1] coordinates {(2.5,0) (3.25,0.15) (4,0)};
    \draw [linecolor] plot [smooth, tension = 1] coordinates {(0.5-2,0) (1.25-2,0.15) (0,0)};
    \draw [linecolor] plot [smooth, tension = 1] coordinates {(1.5,0) (2.25,-0.15) (3,0)};
    \draw [linecolor] plot [smooth, tension = 1] coordinates {(-0.5,0) (0.25,-0.15) (1,0)};
    \draw [linecolor] plot [smooth, tension = 1] coordinates {(3.5,0) (4.25,-0.15) (5,0)};

    \foreach \x in {0,1,2,3} {
        \fill[black, anchor = north] (\x,0) circle (1pt);
    }
    
    \foreach \x in {0,1,2,3} {
        \fill[hcolor, anchor = south] (\x+0.5-2*\squarewidth,+\squarewidth) rectangle (\x+0.5+2*\squarewidth,-\squarewidth);
    }
    
    Place main points and labels
    \def\hshift{0}
    
    \node[vertexcolor,anchor = south, scale = \scale] at (0+\hshift,0) {$e_m$};
    \node[vertexcolor,anchor = north, scale = \scale] at (1+\hshift,0) {$x$};
    \node[vertexcolor,anchor = south, scale = \scale] at (2+\hshift,0) {$x^2$};
    \node[vertexcolor,anchor = north, scale = \scale] at (3+\hshift,0) {$x^3$};

    \def\hshift{0.5}
    \node[anchor = south, scale=\scale, hcolor] at (0+\hshift,0) {$e_m$};
    \node[anchor = north, scale=\scale, hcolor] at (1+\hshift,0) {$x$};
    \node[anchor = south, scale=\scale, hcolor] at (2+\hshift,0) {$x^2$};
    \node[anchor = north, scale=\scale, hcolor] at (3+\hshift,0) {$x^3$};

\end{tikzpicture}
    \caption{Chain complex associated with the cyclic code with check polynomial $e_l+x+x^2$. When the check polynomial is not $p_r(x)$, the visualisation is no longer a circle, although all cells can still be labelled by group algebra elements, and laid out in the same way as in \cref{fig:classical repn code}. Similarly, generalising from the toric code to hypergraph or balanced products of cyclic codes, the visualisation is no longer that of a torus, but all cells are still labelled by group algebra elements from $\mathbb{F}_2C_l\otimes C_m$.}
    \label{fig:1+x+x2 check polynomial}
\end{figure}
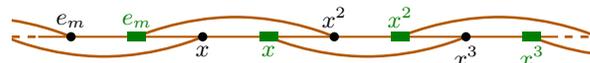

The check polynomials dictate how the $\operatorname{X}$- and $\operatorname{Z}$- checks connect to qubits. 
The repetition code check polynomials $e_m+x$ and $e_l+y$ give a cellulation of the torus with $\operatorname{Z}$-checks as square faces, $\operatorname{X}$-checks as vertices, and qubits as vertical and horizontal edges. Each face is bounded by 4 edges, each edge is connected to 2 vertices.
In the hypergraph product of general cyclic codes, this visualisation is no longer effective. However, the checks and qubits can always be arranged as in \cref{fig:toric layout} on a torus by property of the cyclic group algebra. To illustrate this more clearly, we can consider the chain complex for the classical cyclic code with check polynomial $e_l+x+x^2$, shown in \cref{fig:1+x+x2 check polynomial}. Unlike the chain for the repetition code, this can no longer be interpreted as a circle, as each edge is connected to 3 vertices. However, all edges and vertices are still arranged in the same way.
Therefore for the rest of the paper we will continue referring to qubits as \textit{vertical} and \textit{horizontal} \textit{edges}, the $\operatorname{X}$-checks as \textit{vertices}, and the $Z$-checks as \textit{faces}.

We can refer to $X/Z$ checks and horizontal/vertical qubits by their group algebra element $g\otimes h, g \in C_l, h\in C_m$. Equivalently, with reference to the layout in \cref{fig:toric layout}, we can also refer to them by their \textit{tuple} $(g,h)$. For example, a column of vertical edges indexed by the $e_m$ horizontal translation can be referred to by their row vector $[g_r(y)\otimes e_m|0]$, by $*\otimes e_m$ using group algebra notation, or by their tuple $(*,e_m)$, where the $*$ refers to all elements.

\subsection{Logical Operators from K\"{u}nneth's Formula}
K\"{u}nneth's formula describes an isomorphism from the homology groups of the double complex to the homology group of the total complex; a proof for tensor products is given in~\cite{massey2019basic} and more generally for balanced products in~\cite{Breuckmann_2021_2}. 
We state the formula for tensor products as
\begin{equation}
    \mathcal{H}_n(A \otimes B) \cong \bigoplus_{p+q=n} \mathcal{H}_p(A) \otimes \mathcal{H}_q(B),
\end{equation}
where $A,B$ are the input chains to the double complex, and $A\otimes B$ is the total complex taking the $n^{th}$ term as $\bigoplus_{p+q=n} A_p\otimes B_q$.
This allows us to determine a basis and the number of logical operators for quantum codes given the homology groups $\mathcal{H}_0$ and $\mathcal{H}_1$ of the classical codes in the double complex. 
For repetition codes, the homology group dimensions are given as
\begin{subequations}\label{eq: homology group dimensions}
   \begin{align}
   \begin{split}
       \dim \mathcal{H}_1 &= n- \deg g_r(a)\\
        &= \deg p_r(a) = 1,
    \end{split}\\
    \begin{split}
        \dim \mathcal{H}_0 &= n - (n-\deg p_r(a))\\
        &= \deg p_r(a) = 1,
    \end{split}
    \end{align} 
\end{subequations}
where $n$ is the group order.
The expression for $\dim(\mathcal{H}_1)$ follows from well-known properties of cyclic codes \cite[Ch.~7.\ $\S$~3.\ Theorem 1]{bok:MW}.
Calculation of $\dim(\mathcal{H}_0)$ is less obvious. 
By working with the regular representation, group algebra elements are mapped to $n\times n$ square matrices. 
Then $\mathcal{H}_0=\mathbb{F}_2^{n}/\operatorname{im} p(x)$; and $\dim\operatorname{im} p(x)$ is obtained from considering the dual code, which is also a cyclic code.

We choose representative elements from $\mathcal{H}_1$ and $\mathcal{H}_0$ to be $g_r(a)$ and $e_{|a|}$ respectively. We give an explicit basis of $\operatorname{Z}$-logical operators (in row vector form) for the quantum code as
\begin{equation}
\renewcommand\arraystretch{1.2}
\label{eqn:kun}
    \left(\begin{array}{c|c}
        \begin{matrix}
            g_r(y)\otimes e_m \\[2pt]
        \end{matrix} & 0 \\\hline
         0 & 
         \begin{matrix}
             e_l\otimes g_r(x) \\
         \end{matrix}
    \end{array}\right) = \left(\begin{array}{c}
    \bar{\operatorname{Z}}^v_1 \\\hline
    \bar{\operatorname{Z}}^h_1\end{array}\right) .
\end{equation}
The logical operators are labelled with the superscript $v$ and $h$ representing the way they loop around the handle. The handle supports 2 qubits; this is represented by having one vertical logical and one horizontal logical operator. 
For the toric code, all monomials can be reached through addition of stabilisers, so e.g. the vector $[g_r(y)\otimes x^2 \mid 0]$ is equal to $\bar{\operatorname{Z}}^v_1+\partial_2[g_r(y)\otimes(e_m+x)]$.

\begin{figure}[h!t]
    \centering
\begin{tikzpicture}[scale=2.4, line width = 1pt]
    \colorlet{vertexcolor}{black} %
    \colorlet{linecolor}{orange!70!black}
    \colorlet{hcolor}{green!50!black} %
    \colorlet{vcolor}{blue!80!black}
    \colorlet{xvlog}{yellow!80!black}
    \colorlet{zvlog}{teal!80!black}
    \def\highlightwidth{2}
    \def\highlightopacity{0.5}
    \def\dashstart{-0.5}
    \def\dashend{-0.25}
    \def\linestart{-0.25}
    \def\lineend{2.7}
    \def\scale{1}
    \def\squarewidth{0.035}
    \def\lightwidth{1.8*\squarewidth}
    \foreach \x in {0,1,2} {
        \draw[thick, linecolor] (\x,\linestart) -- (\x,\lineend);
        \draw[dashed, linecolor] (\x,\dashstart) -- (\x,\dashend);
        \draw[dashed, linecolor] (\x,2.75) -- (\x,3);
    }
    
    \foreach \y in {0,1,2} {
        \draw[thick, linecolor] (\linestart,\y) -- (\lineend,\y);
        \draw[dashed, linecolor] (\dashstart,\y) -- (\dashend,\y);
        \draw[dashed, linecolor] (2.75,\y) -- (3,\y);
    }
    
    \foreach \x in {0,1,2} {
        \foreach \y in {0,1,2} {
            \fill[black] (\x,\y) circle (1pt);
        }
    }
    
    \foreach \x in {0,1,2} {
        \foreach \y in {0,1,2} {
            \fill[hcolor] (\x+0.5-2*\squarewidth,\y+\squarewidth) rectangle (\x+0.5+2*\squarewidth,\y-\squarewidth);
        }
    }

    \foreach \x in {0,1,2} {
        \foreach \y in {0,1,2} {
            \fill[vcolor] (\x-\squarewidth,\y+0.5+2*\squarewidth) rectangle (\x+\squarewidth,\y+0.5-2*\squarewidth);
        }
    }

    \foreach \x in {0,1,2} {
        \fill[xvlog, opacity=\highlightopacity] (0.5-2*\lightwidth,\x+\lightwidth) rectangle (0.5+2*\lightwidth,\x-\lightwidth);
    }

    \foreach \x in {0,1,2} {
        \fill[zvlog, opacity = \highlightopacity] (-\lightwidth,\x+0.5+2*\lightwidth) rectangle (\lightwidth,\x+0.5-2*\lightwidth);
    }
    
    \def\hshift{0}
    \def\vshift{-0}
    
    \node[vertexcolor,anchor = north east, scale = \scale] at (0+\hshift,2+\vshift) {$(x^2,e_m)$};
    \node[vertexcolor,anchor = north east, scale = \scale] at (1+\hshift,2+\vshift) {$(x^2,y)$};
    \node[vertexcolor,anchor = north east, scale = \scale] at (2+\hshift,2+\vshift) {$(x^2,y^2)$};
    
    \node[vertexcolor,anchor = north east, scale = \scale] at (0+\hshift,1+\vshift) {$(x,e_m)$};
    \node[vertexcolor,anchor = north east, scale = \scale] at (1+\hshift,1+\vshift) {$(x,y)$};
    \node[vertexcolor,anchor = north east, scale = \scale] at (2+\hshift,1+\vshift) {$(x,y^2)$};
    
    \node[vertexcolor,anchor = north east, scale = \scale] at (0+\hshift,0+\vshift) {$(e_l,e_m)$};
    \node[vertexcolor,anchor = north east, scale = \scale] at (1+\hshift,0+\vshift) {$(e_l,y)$};
    \node[vertexcolor,anchor = north east, scale = \scale] at (2+\hshift,0+\vshift) {$(e_l,y^2)$};
    
    \def\hshift{0.5}
    \def\vshift{0}
    
    \node[anchor = south, scale=\scale, hcolor] at (0+\hshift,2+\vshift) {$(x^2,e_m)$};
    \node[anchor = south, scale=\scale, hcolor] at (1+\hshift,2+\vshift) {$(x^2,y)$};
    \node[anchor = south, scale=\scale, hcolor] at (2+\hshift,2+\vshift) {$(x^2,y^2)$};
    
    \node[anchor = south, scale=\scale, hcolor] at (0+\hshift,1+\vshift) {$(x,e_m)$};
    \node[anchor = south, scale=\scale, hcolor] at (1+\hshift,1+\vshift) {$(x,y)$};
    \node[anchor = south, scale=\scale, hcolor] at (2+\hshift,1+\vshift) {$(x,y^2)$};
    
    \node[anchor = south, scale=\scale, hcolor] at (0+\hshift,0+\vshift) {$(e_l,e_m)$};
    \node[anchor = south, scale=\scale, hcolor] at (1+\hshift,0+\vshift) {$(e_l,y)$};
    \node[anchor = south, scale=\scale, hcolor] at (2+\hshift,0+\vshift) {$(e_l,y^2)$};
    \def\hshift{0}
    \def\vshift{0.5}
    \node[anchor = west, scale=\scale, vcolor] at (0+\hshift,2+\vshift) {$(x^2,e_m)$};
    \node[anchor = west, scale=\scale, vcolor] at (1+\hshift,2+\vshift) {$(x^2,y)$};
    \node[anchor = west, scale=\scale, vcolor] at (2+\hshift,2+\vshift) {$(x^2,y^2)$};
    
    \node[anchor = west, scale=\scale, vcolor] at (0+\hshift,1+\vshift) {$(x,e_m)$};
    \node[anchor = west, scale=\scale, vcolor] at (1+\hshift,1+\vshift) {$(x,y)$};
    \node[anchor = west, scale=\scale, vcolor] at (2+\hshift,1+\vshift) {$(x,y^2)$};
    
    \node[anchor = west, scale=\scale, vcolor] at (0+\hshift,0+\vshift) {$(e_l,e_m)$};
    \node[anchor = west, scale=\scale, vcolor] at (1+\hshift,0+\vshift) {$(e_l,y)$};
    \node[anchor = west, scale=\scale, vcolor] at (2+\hshift,0+\vshift) {$(e_l,y^2)$};
\end{tikzpicture}
    \caption{Toric code layout, as in \cref{fig:toric layout}, but with example $\bar{\operatorname{X}}(\bar{\operatorname{Z}})$ logical operators highlighted in yellow (teal). The $\bar{\operatorname{X}}$ and $\bar{\operatorname{Z}}$ logical operators corresponds to the horizontal and vertical edge sets indexed by $[0 \mid g_r(y)\otimes e_m]$ and $[g_r(y)\otimes e_m \mid 0]$ respectively, and are exactly $\bar{\operatorname{X}}^v_1$ from \cref{eqn:kunx} and $\bar{\operatorname{Z}}^v_1$ from \cref{eqn:kun}. When the meaning is clear, we can refer to these edges simply by their tuple, as in the figure, by their group algebra element, or by their row vector.}
    \label{fig: toric code logical operators}
\end{figure}
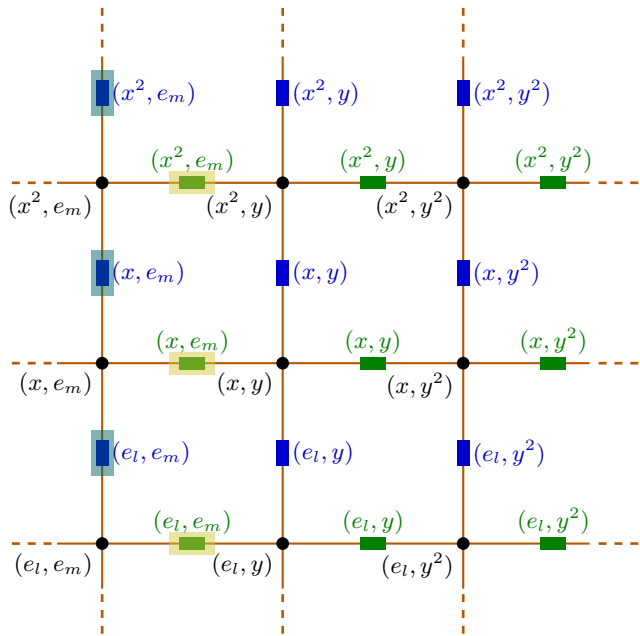

To obtain $\operatorname{X}$-logical operators, we consider the cochain. The coboundary operators are obtained by taking the transpose of the boundary operators, and are given as
\begin{subequations} \label{eqn:coboundary}
\renewcommand\arraystretch{1.2}
\begin{align}
      \delta_1 &= \left[e_l\otimes p_r^T(x)\mid p_r^T(y)\otimes e_m\right],\\
      \delta_0 &=\left[ 
        \begin{array}{c}
            p_r^T(y)\otimes e_m\\\hline
             e_l\otimes p_r^T(x)
        \end{array}
        \right]
\end{align}
\end{subequations}
The notation $f^T(a)$ for polynomial $f$ refers to an element-wise inverse polynomial, where the antipodal map $a\mapsto\bar{a}=a^{-1}$ is applied to every element. The cochain is isomorphic to the base chain, and so we can express the cohomology elements in the same basis of group algebra elements. Therefore to write the $\bar{\operatorname{X}}$ logical operators, we only have to define the element-wise inverse polynomials $p_r^T(x)$ and $g_r^T(x)$, which are associated with the classical code defined by the transposed parity check matrices. For the toric code, the transposed and non-transposed codes are the same because $p_r^T(x) = x^{-1}p_r(x)$. Therefore we can re-use the basis in \cref{eqn:kun} and write an X-logical basis as
\begin{equation}
\renewcommand\arraystretch{1.2}
    \label{eqn:kunx}
    \left(\begin{array}{c|c}
       
        0 &
         \begin{matrix}
            g_r(y)\otimes e_m
        \end{matrix} \\\hline
         \begin{matrix}
            e_l\otimes g_r(x) \\
        \end{matrix} & 0
    \end{array}\right) = \left(\begin{array}{c}
    \bar{\operatorname{X}}^v_1 \\\hline
    \bar{\operatorname{X}}^h_1\end{array}\right).
\end{equation}
An illustration of the logical operators is given in \cref{fig: toric code logical operators}.
\subsection{Dehn Twists on the Toric Code}
A Dehn twist is implemented on the toric code by taking an entire column (row) of vertical (horizontal) edges, and applying $\operatorname{CNOT}$ gates sequentially, first to the adjacent horizontal (vertical) edge, then to the one translated one step vertically (horizontally), repeating with increasing vertical (horizontal) translations until all edges in the loop have been targeted.

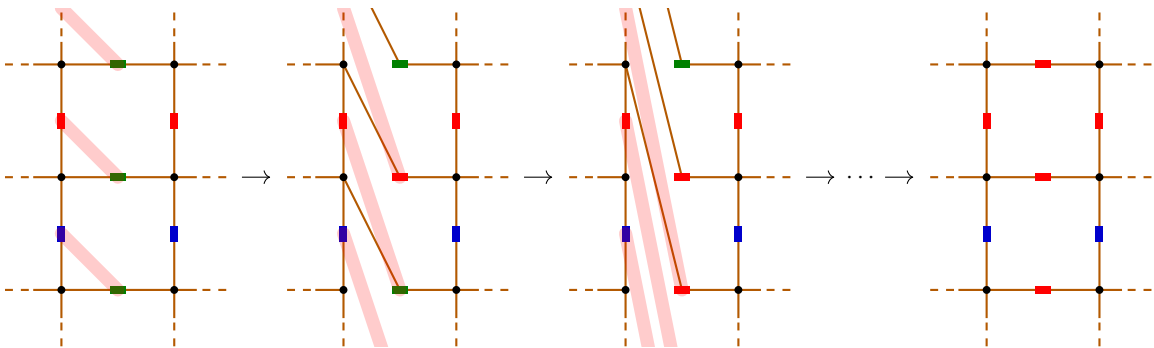
\begin{figure*}
    \centering
\begin{tikzpicture}[scale=1.5]
    \colorlet{vertexcolor}{black} %
    \colorlet{linecolor}{orange!70!black}
    \colorlet{hcolor}{green!50!black} %
    \colorlet{vcolor}{blue!80!black}
    \colorlet{xvlog}{yellow!80!black}
    \colorlet{zvlog}{teal!80!black}
    \colorlet{highlightcolor}{teal!70!black}
    \colorlet{overlapcolor}{red}
    \def\highlightwidth{5}
    \def\highlightopacity{0.2}
    \def\dashstart{-0.5}
    \def\dashend{-0.25}
    \def\linestart{-0.25}
    \def\lineend{1.2}
    \def\scale{1}
    \def\arrowlength{0.25}
    \def\squarewidth{0.035}
    \def\lightwidth{1.8*\squarewidth}
    \def\trianglewidth{0.1}
    \def\root{1.73205080757}

\begin{scope}[xshift=0cm]
  \clip (\dashstart,\dashstart) rectangle (1.5,2.5);
    \foreach \x in {0,1} {
        \draw[thick, linecolor] (\x,\linestart) -- (\x,2.2);
        \draw[thick, dashed, linecolor] (\x,\dashstart) -- (\x,\dashend);
        \draw[thick, dashed, linecolor] (\x,2.25) -- (\x,2.5);
    }
    
    \foreach \y in {0,1,2} {
        \draw[thick, linecolor] (\linestart,\y) -- (\lineend,\y);
        \draw[thick, dashed, linecolor] (\dashstart,\y) -- (\dashend,\y);
        \draw[thick, dashed, linecolor] (1.25,\y) -- (1.5,\y);
    }
    
    \foreach \x in {0,1} {
        \foreach \y in {0,1,2} {
            \fill[black] (\x,\y) circle (1pt);
        }
    }
    
    \foreach \x in {0} {
        \foreach \y in {0,1,2} {
            \fill[hcolor] (\x+0.5-2*\squarewidth,\y+\squarewidth) rectangle (\x+0.5+2*\squarewidth,\y-\squarewidth);
        }
    }
    \foreach \x in {0,1} {
        \foreach \y in {0,1} {
            \fill[vcolor] (\x-\squarewidth,\y+0.5+2*\squarewidth) rectangle (\x+\squarewidth,\y+0.5-2*\squarewidth);
        }
    }

    \fill[overlapcolor] (-\squarewidth,1.5+2*\squarewidth) rectangle (+\squarewidth,1.5-2*\squarewidth);

    \fill[overlapcolor] (1-\squarewidth,1.5+2*\squarewidth) rectangle (1+\squarewidth,1.5-2*\squarewidth);

    \foreach \y in {0,1,2} {
        \draw[red, line width = \highlightwidth pt, opacity = \highlightopacity, line cap = round, scale = \scale] (0,\y+0.5) -- (0.5, \y);
    }

\end{scope}

\def\arrowstart{1.6}
\draw[->] (\arrowstart,1) -- (\arrowstart+\arrowlength,1);

\begin{scope}[xshift=2.5cm]
\clip (\dashstart,\dashstart) rectangle (1.5,2.5);
    \foreach \x in {0,1} {
        \draw[thick, linecolor] (\x,\linestart) -- (\x,2.2);
        \draw[thick, dashed, linecolor] (\x,\dashstart) -- (\x,\dashend);
        \draw[thick, dashed, linecolor] (\x,2.25) -- (\x,2.5);
    }
    
    \foreach \y in {0,1,2} {
        \draw[thick, linecolor] (\linestart,\y) -- (0,\y);
        \draw[thick, linecolor] (0.5,\y) -- (\lineend,\y);
        \draw[thick, linecolor] (0,\y+1) -- (0.5,\y);
        \draw[thick, dashed, linecolor] (\dashstart,\y) -- (\dashend,\y);
        \draw[thick, dashed, linecolor] (1.25,\y) -- (1.5,\y);
    }
    
    \foreach \x in {0,} {
        \foreach \y in {0,1,2} {
            \fill[black] (\x,\y) circle (1pt);
        }
    }
    
    \foreach \x in {0} {
        \foreach \y in {0,1,2} {
            \fill[hcolor] (\x+0.5-2*\squarewidth,\y+\squarewidth) rectangle (\x+0.5+2*\squarewidth,\y-\squarewidth);
        }
    }

    \foreach \x in {0,1} {
        \foreach \y in {0,1} {
            \fill[vcolor] (\x-\squarewidth,\y+0.5+2*\squarewidth) rectangle (\x+\squarewidth,\y+0.5-2*\squarewidth);
        }
    }

    \fill[overlapcolor] (0.5-2*\squarewidth,1+\squarewidth) rectangle (0.5+2*\squarewidth,1-\squarewidth);
    \fill[overlapcolor] (-\squarewidth,1.5+2*\squarewidth) rectangle (\squarewidth,1.5-2*\squarewidth);
    \fill[overlapcolor] (1-\squarewidth,1.5+2*\squarewidth) rectangle (1+\squarewidth,1.5-2*\squarewidth);
    \foreach \y in {0,1,2} {
        \draw[red, line width = \highlightwidth pt, opacity = \highlightopacity, line cap = round, scale = \scale] (0,\y+0.5) -- (0.5, \y-1);
    }
\end{scope}

\def\arrowstart{4.1}
\draw[->] (\arrowstart,1) -- (\arrowstart+\arrowlength,1);

\begin{scope}[xshift=5cm]
    \clip (\dashstart,\dashstart) rectangle (1.5,2.5);
    \foreach \x in {0,1} {
        \draw[thick, linecolor] (\x,\linestart) -- (\x,2.2);
        \draw[thick, dashed, linecolor] (\x,\dashstart) -- (\x,\dashend);
        \draw[thick, dashed, linecolor] (\x,2.25) -- (\x,2.5);
    }
    
    \foreach \y in {0,1,2} {
        \draw[thick, linecolor] (\linestart,\y) -- (0,\y);
        \draw[thick, linecolor] (0.5,\y) -- (\lineend,\y);
        \draw[thick, linecolor] (0,\y+2) -- (0.5,\y);
        \draw[thick, dashed, linecolor] (\dashstart,\y) -- (\dashend,\y);
        \draw[thick, dashed, linecolor] (1.25,\y) -- (1.5,\y);
    }
    
    \foreach \x in {0,} {
        \foreach \y in {0,1,2} {
            \fill[black] (\x,\y) circle (1pt);
        }
    }
    
    \foreach \x in {0} {
        \foreach \y in {0,1,2} {
            \fill[hcolor] (\x+0.5-2*\squarewidth,\y+\squarewidth) rectangle (\x+0.5+2*\squarewidth,\y-\squarewidth);
        }
    }

    \foreach \x in {0,1} {
        \foreach \y in {0,1} {
            \fill[vcolor] (\x-\squarewidth,\y+0.5+2*\squarewidth) rectangle (\x+\squarewidth,\y+0.5-2*\squarewidth);
        }
    }

    \fill[overlapcolor] (0.5-2*\squarewidth,1+\squarewidth) rectangle (0.5+2*\squarewidth,1-\squarewidth);
    \fill[overlapcolor] (0.5-2*\squarewidth,\squarewidth) rectangle (0.5+2*\squarewidth,-\squarewidth);
    \fill[overlapcolor] (-\squarewidth,1.5+2*\squarewidth) rectangle (\squarewidth,1.5-2*\squarewidth);
    \fill[overlapcolor] (1-\squarewidth,1.5+2*\squarewidth) rectangle (1+\squarewidth,1.5-2*\squarewidth);
    \foreach \y in {0,1,2} {
        \draw[red, line width = \highlightwidth pt, opacity = \highlightopacity, line cap = round, scale = \scale] (0,\y+0.5) -- (0.5, \y-2);
    }
\end{scope}

\def\arrowstart{6.6}
\draw[->] (\arrowstart,1) -- (\arrowstart+\arrowlength,1);

\node at (\arrowstart+\arrowlength+0.25,1) {$\cdots$};

\def\shift{8.2}
\def\arrowstart{\shift-0.9}
\draw[->] (\arrowstart,1) -- (\arrowstart+\arrowlength,1);

\begin{scope}[xshift=\shift cm]
  \clip (\dashstart,\dashstart) rectangle (1.5,2.5);
    \foreach \x in {0,1} {
        \draw[thick, linecolor] (\x,\linestart) -- (\x,2.2);
        \draw[thick, dashed, linecolor] (\x,\dashstart) -- (\x,\dashend);
        \draw[thick, dashed, linecolor] (\x,2.25) -- (\x,2.5);
    }
    
    \foreach \y in {0,1,2} {
        \draw[thick, linecolor] (\linestart,\y) -- (\lineend,\y);
        \draw[thick, dashed, linecolor] (\dashstart,\y) -- (\dashend,\y);
        \draw[thick, dashed, linecolor] (1.25,\y) -- (1.5,\y);
    }
    
    \foreach \x in {0,1} {
        \foreach \y in {0,1,2} {
            \fill[black] (\x,\y) circle (1pt);
        }
    }
    
    \foreach \x in {0} {
        \foreach \y in {0,1,2} {
            \fill[overlapcolor] (\x+0.5-2*\squarewidth,\y+\squarewidth) rectangle (\x+0.5+2*\squarewidth,\y-\squarewidth);
        }
    }

    \foreach \x in {0,1} {
        \foreach \y in {0,1} {
            \fill[vcolor] (\x-\squarewidth,\y+0.5+2*\squarewidth) rectangle (\x+\squarewidth,\y+0.5-2*\squarewidth);
        }
    }

    \fill[overlapcolor] (-\squarewidth,1.5+2*\squarewidth) rectangle (\squarewidth,1.5-2*\squarewidth);
    \fill[overlapcolor] (1-\squarewidth,1.5+2*\squarewidth) rectangle (1+\squarewidth,1.5-2*\squarewidth);
\end{scope}

\end{tikzpicture}
    \caption{The first three steps of a Dehn twist, and the final result of a Dehn twist on the toric code. The layout is the same as in \cref{fig:toric layout} but with vertex labels omitted. The twist applies $\operatorname{CNOT}$ gates, represented as red highlights, from a column of vertical edges (blue vertical rectangles) to a column of horizontal edges (green horizontal rectangles). For an $l\times m$ toric code, we require $l$ rounds of $\operatorname{CNOT}$ gates to implement the full twist; correspondingly a row twist scheme would require $m$ rounds of $\operatorname{CNOT}$ gates. The twist overlaps on 1 edge modulo 2 of $\bar{\operatorname{X}}^h$, colored red, and propagates the Pauli $\operatorname{X}$ from this edge to horizontal edges. After the full twist, the overlap has propagated to a vertical logical $\bar{\operatorname{X}}^v$, and we have returned to the original toric code. This shows that the Dehn twist acts as a logical $\operatorname{CNOT}$ gate.}
    \label{fig: dehn twist}
\end{figure*}

In the following discussions, we consider implementing a vertical twist onto an $l\times m$ toric code. A schematic of this is given in \cref{fig: dehn twist}. The twist gives the map (ignoring identity mappings)
\begin{subequations} \label{eqn:dehn twist mapping}
    \begin{align}
        \bar{\operatorname{X}}^h_1 &\mapsto \bar{\operatorname{X}}^h_1\bar{\operatorname{X}}^v_1\\
        \bar{\operatorname{Z}}^h_1 &\mapsto \bar{\operatorname{Z}}^h_1\bar{\operatorname{Z}}^v_1
    \end{align}
\end{subequations}
A row twist would implement the same mapping but with $h$ and $v$ superscripts exchanged. The vertical twist adds a vertical essential cycle to the horizontal one, which is guaranteed to overlap on 1 edge modulo 2 with the column. Boundary and coboundary operators remain the same after the twist, returning us the same quantum code. Dehn twists act as logical $\operatorname{CNOT}$ gates between the qubit supported on the torus vertically and horizontally.

In the following sections, we explain how the schematic of \cref{fig: dehn twist} implements the mapping in \cref{eqn:dehn twist mapping}. We first describe how $\operatorname{CNOT}$ operators affect Pauli operators through conjugation, and in turn the boundary and coboundary operators. Using this, we show that the Dehn twist propagates Pauli operators such that one essential cycle is added to the other. Finally, we show that the boundary and coboundary operators do not change after the twist, which returns us the same quantum code.
\subsubsection{Action of $\operatorname{CNOT}$ operators on Pauli operators}
The conjugation relations of the $\operatorname{CNOT}$ gate with Pauli $\operatorname{X}$ and $\operatorname{Z}$ are given by
\begin{subequations} \label{eqn:cnot action}
    \begin{align}
        \operatorname{CNOT}_{1,2} X_1 \operatorname{CNOT}_{1,2}^{-1} &= X_1X_2,\\
        \operatorname{CNOT}_{1,2} Z_2 \operatorname{CNOT}_{1,2}^{-1} &= Z_1Z_2,
    \end{align}
\end{subequations}
where the subscripts index some qubit on which the operators act on, and $\operatorname{CNOT}_{i,j}$ acts from control qubit $i$ to target qubit $j$.

Using these relations, we can determine how the boundary operators change when $\operatorname{CNOT}$ gates are applied between control and target qubits.
The operator $\delta_0$ %
indicates the qubits connected to each check. By \cref{eqn:cnot action}, applying a $\operatorname{CNOT}$ gate from a control qubit connected to a check \textit{adds} the target qubit to $\delta_0$ limited to the check. 
On \cref{fig: dehn twist}, this is represented by the addition of connections between the target qubit and $\operatorname{X}$-checks connected to the control qubit. 
Working in~$\mathbb{F}_2$ means that connections already present are broken. Visually, this shows how the Dehn twist sequentially cuts and reconnects edges, akin to twisting by $2\pi$ on a torus.

By the same \cref{eqn:cnot action}, the $\operatorname{CNOT}$ gate also propagates any Pauli $\operatorname{X}$ from the control qubit to the target qubit. Therefore after a full Dehn twist, the Pauli $\operatorname{X}$ on the overlapping edge of $\bar{\operatorname{X}}^h$ is propagated to $\bar{X}^v$. This is the mechanism by which the Dehn twist propagates logical operators conditioned on overlap with another logical operator.

Symmetrically, the action of $\operatorname{CNOT}$ on Pauli $\operatorname{Z}$ adds a transpose term to $\operatorname{Z}$-checks in $\partial_2$ and also propagates Pauli $\operatorname{Z}$ operators from the target to control qubit. We can therefore construct the boundary and coboundary operators after $\operatorname{CNOT}$ gates solely by considering their effect on $\delta_0$. 
Explicitly,
\begin{subequations} \label{eqn:operators during twist}
\renewcommand\arraystretch{1.2}
    \begin{align}
      \partial_1 &= \left[A\mid B\right],\\
      \partial_2 &=\left[ 
        \begin{array}{c}
            B\\\hline
             A
        \end{array}
        \right],\\
      \delta_1 &= \left[B^T\mid A^T\right],\\
      \delta_0 &=\left[ 
        \begin{array}{c}
            A^T\\\hline
             B^T
        \end{array}
        \right]
    \end{align}
\end{subequations}
for some $A,B$ in the tensor product group algebra, where the superscript $T$ indicates the transpose, restricted to the $\operatorname{X}$-checks that are connected to the control qubit. We can infer the form of other boundary and coboundary operators from $\delta_0$. Therefore, for further discussions, we mostly consider $\delta_0$ and the action of the Dehn twist on $\bar{\operatorname{X}}$, and leave implicit the construction of other coboundary and boundary operators using \cref{eqn:operators during twist}.
\subsubsection{Pauli propagation}
Consider an example where we twist on the column indexed by the horizontal translation $x^k$, i.e. the set of edges and vertices in the toric layout of \cref{fig:toric layout} given by the tuple $(*,x^k)$. Then the $\operatorname{CNOT}$ gates will be from vertical edge set $[g_r(y)\otimes x^k \mid 0]$ to horizontal edge set $[0 \mid g_r(y)\otimes x^k]$. 

In general, the overlap of $\bar{\operatorname{X}}^h$ with the twist can occur on some vertical edge set $[f(y)\otimes x^k \mid 0]$, which is the overlap of the twist with $\bar{\operatorname{X}}^h_1$ plus stabilisers. 
Algebraically, the $i^{th}$ round of $\operatorname{CNOT}$ gates as shown in \cref{fig: dehn twist} can be thought of as propagating the Pauli $\operatorname{X}$ on this overlap to the horizontal edge set $[0\mid y^{-i+1} f(y) \otimes x^k]$. 

Symmetrically, let the overlap of $\bar{\operatorname{Z}}^h$ with the twist occur on an horizontal edge set ${[0 \mid f'(y)\otimes x^k]}$. 
In the $i^{th}$ round of $\operatorname{CNOT}$ gates, the twist would propagate the Pauli $\operatorname{Z}$ from this overlap to the vertical edge set $[y^{i-1} f'(y)\otimes x^k|0]$. 
After $l$ rounds of $\operatorname{CNOT}$ gates, the original overlapping Pauli $\operatorname{X}$ and $\operatorname{Z}$ would have propagated to the edges ${[g_r^T(y)f(y)\otimes x^k \mid 0]}$ and ${[0\mid g_r(y)f'(y)\otimes x^k]}$ respectively. 
Note that the same scheme implements $g^T(y)$ on the $\operatorname{X}$-logical operators but~$g(y)$ on the $\operatorname{Z}$-logical operators due to \cref{eqn:cnot action}.

For the toric code, $g_r^T(y)=g_r(y)$ since $g_r(y)$ is a linear sum of \textit{every} $g\in C_l$. This is an all-one vector, and taking the transpose polynomial has the effect of permuting the ones in the vector, which leaves the vector invariant. We are also guaranteed that the polynomials $f(y),f'(y)$ have weight 1 modulo 2. Moreover all monomials in the toric code are in the same equivalence class, and can be reached by addition of stabilisers. Therefore, the propagated edges can be translated down to $\bar{\operatorname{X}}_1^v$ and $\bar{\operatorname{Z}}^v_1$, which is exactly the mapping described in \cref{eqn:dehn twist mapping}.
\subsubsection{Action on the coboundary operators}
Consider the action of the twist restricted only to the column of $\operatorname{X}$-checks indexed by $g_r(y)\otimes x^k$. We show that after the twist, $\delta_0$ (and by extension the other boundary and coboundary operators) remains invariant.

For each $\operatorname{X}$-check indexed by $h\otimes x^k,h\in C_l$, the $i^{th}$ $\operatorname{CNOT}$ round acts from all connected vertical edges ${[p^T_r(y)h\otimes x^k\mid 0]}$ and adds associated horizontal edges ${[0\mid p^T_r(y)hy^{-i+1}\otimes x^k]}$ to the coboundary operator. 
Therefore the local coboundary operator of the $\operatorname{X}$-checks restricted to the twist column $(*,x^k)$ after $i$ $\operatorname{CNOT}$ rounds is given by
\begin{equation} \label{eqn:coboundary_lround}
\renewcommand\arraystretch{1.2}
    \delta_0^{(i)}|_{twist} =\left[ 
        \begin{array}{c}
            p_r^T(y)\otimes e_m\\\hline
             e_l\otimes p_r^T(x) + p^T_r(y)(y^{0}+\dots+y^{-i+1})\otimes e_m
        \end{array}
        \right].
\end{equation}
For a concrete example, after one round of $\operatorname{CNOT}$ gates, the local coboundary operator on column $(*,x^k)$ for the toric code looks like
\begin{equation}
    \delta_0^{(1)}|_{twist} =\left[ 
        \begin{array}{c}
            (e_l+\bar{y})\otimes e_m\\\hline
             e_l\otimes \bar{x}+\bar{y}\otimes e_m
        \end{array}\right]
\end{equation}
from the addition of $p_r^T(y)e_l\otimes e_m = (e_l+\bar{y})\otimes e_m$ horizontal edges. Note here that by applying the twist from \textit{all} edges in the column, we guarantee that every $\operatorname{X}$-check will have $p_r^T(y)$ in the additional coboundary operator term. Comparing this to the original coboundary operator, we see that the horizontal edge term $e_l\otimes e_m$, which attached the vertex to the right horizontal edge as described in \cref{fig:toric layout}, has been cancelled and replaced with $\bar{y}\otimes e_m$. This connects the vertex to the right edge translating a step down like in \cref{fig: dehn twist}. Further steps in the twist cause this cancellation and addition of further vertically-translated right edges sequentially.

Completing the full twist causes the additional term to sum as
\begin{equation} \label{eqn:toric code single step}
\begin{split}
    \left(p^T_r(y)\otimes e_m\right)\left(\sum_{i=1}^{l}y^{-i+1}\otimes e_m\right) &= p^T_r(y)g_r^T(y)\otimes e_m\\
    &=0
\end{split}
\end{equation}
due to the classical code property 
\begin{subequations} \label{eqn: vanishing add term}
    \begin{align}
        p_r^T(x)g_r^T(x) &= 0,\\
        p_r(x)g_r(x) &= 0,
    \end{align}
\end{subequations}
which extends \cref{eqn: vanishing add term} to the transpose polynomial straightforwardly.
The action of the $\operatorname{CNOT}$ gates on the boundary operators is given symmetrically by taking the transpose of the coboundary operators; we have therefore algebraically shown how the initial code is recovered: 
at the end of the twist, additional terms in the coboundary operator sum to 0. 
The scheme for horizontal Dehn twists can be shown to work in the same way, except now the additional term occurs in the space of vertical edges. 
However by \cref{eqn: vanishing add term} we will recover the same code as we started with.

In the toric code, there is a visual representation of how the edges are translated by considering \cref{eqn:coboundary_lround,eqn:toric code single step}. In general cyclic codes, this visualisation is no longer useful, and we instead turn to \cref{eqn: vanishing add term} to show we return to the original qLDPC code.
\section{Tensor Product of General Cyclic Codes} \label{sec:general}
In this section, we generalise Dehn twists to the hypergraph and balanced product of cyclic codes, which arise from the same double complex construction of cyclic group algebras. We show that work done in describing the toric code as a chain complex, as well as with polynomials from cyclic group algebras allow us to do this straightforwardly.
\subsection{Logical Basis and Boundary Operators of the Hypergraph Product}
Following the discussion with the toric code, note that if we define $\alpha_p:=\deg p(x)$ for a product of some \textit{irreducible polynomials}\footnote{The distinction is important as, in general, we can take some representative polynomial with higher degree.} $p(x)$ then we can write down a canonical basis for classical cyclic codes as
\begin{subequations} \label{eqn: homology group representatives}
\begin{align}
    \mathcal{H}_1 &= \langle g(x), x\cdot g(x), \dots, x^{\alpha_p-1}g(x)\rangle,\\
    \mathcal{H}_0 &= \langle e, x,\dots, x^{\alpha_p-1}\rangle.
\end{align}
\end{subequations}
The basis of $\mathcal{H}_1$ is from the definition of a cyclic code. Further cyclic shifts of $g(x)$ are no longer linearly-independent. Similarly, for $\mathcal{H}_0$, we take monomials of degree lower than $\alpha_p$, which are not in $\operatorname{im} p(x)$. 

In general we have freedom to choose representative polynomials used in code construction as long as they have $p(x)$ as a factor. %
For the hypergraph product, the choice of the representative polynomial does not matter as the underlying classical code is the same, and therefore the quantum code will have the same parameters, albeit with check weight depending on the weight of the representative polynomials used.

Consider 2 cyclic codes with check polynomials $p_1(y), p_2(x)$ and generator polynomials $g_1(y), g_2(x)$. Generalising from \cref{eqn:boundary}, we can define boundary and coboundary operators of the total complex as
\begin{subequations} \label{eqn:gen_operators}
\renewcommand\arraystretch{1.2}
    \begin{align}
      \partial_1 &= \left[p_1(y)\otimes e_m\mid e_l\otimes p_2(x)\right],\\
      \partial_2 &=\left[ 
        \begin{array}{c}
            e_l\otimes p_2(x)\\\hline
             p_1(y)\otimes e_m
        \end{array}
        \right],\\
      \delta_1 &= \left[e_l\otimes p_2^T(x)\mid p_1^T(y)\otimes e_m\right],\\
      \delta_0 &=\left[ 
        \begin{array}{c}
            p_1^T(y)\otimes e_m\\\hline
             e_l\otimes p_2^T(x)
        \end{array}
        \right],
    \end{align}
\end{subequations}
Following \cref{eqn:kun,eqn:kunx}, we can also write down a basis of logical operators given the homology representatives in \cref{eqn: homology group representatives}. We define
\begin{subequations} \label{eqn: general logical basis}
    \begin{align}
        \bar{\operatorname{Z}}^h_{i,j} &= [0 \mid y^{i-1}\otimes x^{j-1}g_2(x)]\\
        \bar{\operatorname{Z}}^v_{i,j} &= [y^{i-1}g_1(y)\otimes x^{j-1} \mid 0]\\
        \bar{\operatorname{X}}^h_{i,j} &= [y^{i-1}\otimes x^{j-1}g_2^T(x) \mid 0]\\
        \bar{\operatorname{X}}^v_{i,j} &= [0 \mid y^{i-1}g_1^T(y)\otimes x^{j-1}]
    \end{align}
\end{subequations}
where
\begin{subequations}
    \begin{align}
        1\leq i &\leq \alpha_{p_1},\\
        1\leq j &\leq \alpha_{p_2}.   
    \end{align}
\end{subequations}

The notation $\alpha_{p_i}$ refers to the degree of the irreducible polynomial associated with $p_i(a)$. The degrees of the transposed polynomials will be the same as that of the untransposed polynomials\footnote{The degree will be the same but the polynomials themselves might not be. For example, transposing the check polynomial for the $[7,4,3]$ Hamming code gives another Hamming code with the same parameters, but one will have $p(x)=(1+x)(1+x+x^3)$ while the other will have $(1+x)(1+x^2+x^3)$.} as their regular representation matrix ranks are the same, and by application of the rank-nullity theorem.

We refer back to the toric layout of the checks and qubits as in \cref{fig:toric layout}. Then, we can interpret the $y^{i-1}$ and $x^{j-1}$ as vertical and horizontal translations of the logical operators on the two-dimensional grid respectively. This gives us a geometric realisation of the total complex homology and cohomology groups. The subscripts $i,j$ in this notation can be interpreted as maintaining the structure of the double complex. By K\"unneth's formula, the number of logical qubits $k$ is given by
\begin{equation}
    k = 2 \alpha_{p_1} \alpha_{p_2}.
\end{equation}
Here, unlike the toric code, more than 2 logical qubits can be supported per handle. The additional supported qubits mean that Dehn twists can overlap on multiple qubits, which expands the number of possible Dehn twists to implement different logical operators. 
\subsection{Generalised Dehn Twists}
In the toric code, we explicitly showed that additional terms in the boundary and coboundary operators vanish in \cref{eqn:toric code single step} after the Dehn twist is completed.
The idea behind a general version of the Dehn twists is to exploit \cref{eqn: vanishing add term} so that additional terms in the boundary and coboundary operators in the hypergraph and balanced products of other cyclic codes also vanish after the twist is completed.
We can ensure this happens by twisting on \textit{all} edges connected to the checks, to have $\operatorname{CNOT}$ gates from a set of all vertical (horizontal) edges in a column (row) to another set of all horizontal (vertical) edges in a column (row). Twisting on all edges guarantees that the additional coboundary term contains $p_i^T(a)$; implementing the logical will multiply that by $g_i^T(a)$.

As in the previous section on the toric code, we consider a vertical twist. There are two main differences between a generalised twist scheme and that for the toric code. 

Firstly, there is a choice in the columns to twist from and to. With logical operators of the form in \cref{eqn: general logical basis}, there are at least $\alpha_{p_2}$ columns with different overlaps on $\bar{\operatorname{X}}$ and $\bar{\operatorname{Z}}$ logical operators respectively, due to horizontal translations $x^{j-1}$. Moreover, each column overlaps all $i$ vertical translations of logical operators, for which there are $\alpha_{p_1}$ per column. A single twist will have an effect on multiple logical operators simultaneously. Overlapping on the logical operators when implementing the twist will cause every overlapping logical to act as a control.

Secondly, there is also a choice of which vertical logical operators to implement.
Again by property of the cyclic codes, there are at least $\alpha_{p_1}$ different Dehn twists corresponding to the vertical logical operators that we can propagate an edge to.
We obtain the twists by first explicitly identifying an overlapping edge of the twist column with a single logical.
To implement an $\operatorname{X}$-logical operator, from this edge, target the edge set of the vertical logical sequentially with $\operatorname{CNOT}$ gates, and copy this action for all edges in the control qubit column to the target qubit column. For $\operatorname{Z}$-logical operators, we reverse the $\operatorname{CNOT}$ target and control qubits.

The overlap of the twist on other vertically-translated logical operators will then propagate to similarly vertically-translated logical operators. In general we then have to use the classical cyclic code properties to determine how to express them in our given logical basis.

Providing a $\operatorname{CNOT}$ scheme that implements $\operatorname{\bar{X}}$ or $\operatorname{\bar{Z}}$ logical operators implements the transpose polynomial for the other Pauli operator. This is because of the symmetry in the $\operatorname{CNOT}$ action on the Pauli operators as stated in \cref{eqn:cnot action}, where target and control qubits are swapped for $\operatorname{X}$ and $\operatorname{Z}$. In the toric code, $g_r(x)=g_r^T(x)$, but this is not true for general cyclic codes.
In general, the combined effect of a Dehn twist on $\bar{\operatorname{X}}$ and $\bar{\operatorname{Z}}$ has to be found numerically if the twist columns do not fully support all logical $\bar{\operatorname{X}}$ or $\bar{\operatorname{Z}}$ operators.

Finally, note that just as in the toric code, the number of $\operatorname{CNOT}$ rounds required scales as the code distance~$d$. 
We need $d$ rounds of $\operatorname{CNOT}$ gates to propagate the overlapping Pauli to another logical operator of distance $d$.
We also need $\mathcal{O}(d)$ $\operatorname{CNOT}$ gates per round, so the number of physical operations for the Dehn twist scales as $\mathcal{O}(d^2)$. However, the intermediate code is known, there is no additional overhead for this scheme in terms of ancillary qubits, and the time overhead for the scheme is $d$.

\subsection{Instantaneous Twists}
The notation used in writing the coboundary and boundary operators also allow us to consider \textit{instantaneous twists} \cite{Breuckmann_2017} analytically. Roughly speaking, an instantaneous twist reduces the time overhead for implementing a Dehn twist by applying $\operatorname{CNOT}$ gates in parallel on an extended spatial region of the code instead of a single row or column, which can also reduce connectivity requirements.
In the example of the toric code, instantaneous twists can be implemented by applying a single round of $\operatorname{CNOT}$ gates on \textit{every} vertical edge indexed by $(h,g)$ to the equivalent horizontal edge $(h,g)$. This implements the Dehn twist in a single time step. The form of the coboundary operator after the twist is given by
\begin{equation} \label{eqn: toric code inst twist}
    \delta_0^{(1)} =\left[ 
        \begin{array}{c}
            (e_l+\bar{y})\otimes e_m\\\hline
             e_l\otimes \bar{x}+\bar{y}\otimes e_m
        \end{array}\right].
\end{equation}
Note that the coboundary operator is no longer restricted to the column on which we twist, because now the twist is applied everywhere. Relabelling the group algebra elements with an additional vertical translation term returns us the boundary operators of the toric code.

For general cyclic codes, we find that it is not possible to generalise instantaneous twists for this scheme. A single round of $\operatorname{CNOT}$ gates adds an additional $p_1^T(y)g^{(1)}(y)$ term to the boundary operators, where the notation $g^{(k)}(y)$ refers to the first $k$ terms in the polynomial $g(y)$. This cannot be relabelled to return us back to the original code. We see that instantaneous twists can be implemented for the toric code because of the repetition code and its symmetries: the coboundary operator in \cref{eqn: toric code inst twist} has 4 terms corresponding to 4 edges after 1 round of $\operatorname{CNOT}$ gates due to the fact that
\begin{equation}
    p_r(y)g_r^{(k)}(y) = e_l+y^{k+1}.
\end{equation}

\subsection{Visual Example of a Generalised Dehn Twist}
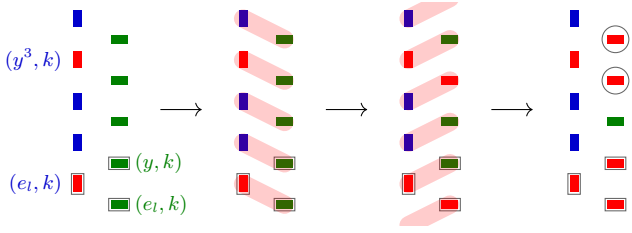
\begin{figure}
    \centering
\begin{tikzpicture}[scale=2.2]
    \colorlet{vertexcolor}{black} %
    \colorlet{linecolor}{orange!70!black}
    \colorlet{hcolor}{green!50!black} %
    \colorlet{vcolor}{blue!80!black}
    \colorlet{xvlog}{yellow!80!black}
    \colorlet{zvlog}{teal!80!black}
    \colorlet{highlightcolor}{teal!70!black}
    \colorlet{overlapcolor}{red}
    \def\circlewidth{2.3}
    \def\circlehighlight{white!40!black} %
    \def\highlightwidth{7}
    \def\highlightopacity{0.2}
    \def\dashstart{-0.5}
    \def\dashend{-0.25}
    \def\linestart{-0.25}
    \def\lineend{1.2}
    \def\scale{1}
    \def\nodescale{0.85}
    \def\arrowlength{0.25}
    \def\arrowvertpos{0.7}
    \def\squarewidth{0.025}
    \def\trianglewidth{0.1}
    \def\root{1.73205080757}
\begin{scope}[xshift=0cm]
  \clip (-.5,0) rectangle (0.75,1.35);

    \def\yshift{-.125}
    \def\xshift{-0.05}
    \node[vcolor,anchor = east, scale = \nodescale] at (0+\xshift,0.25) {$(e_l,k)$};
    \node[vcolor,anchor = east, scale = \nodescale] at (0+\xshift,1) {$(y^3,k)$};

    \node[hcolor,anchor = west, scale = \nodescale] at (0.25-\xshift,.25+\yshift) {$(e_l,k)$};
    \node[hcolor,anchor = west, scale = \nodescale] at (0.25-\xshift,.5+\yshift) {$(y,k)$};
    
    \foreach \y in {.25,.5,.75,1,1.25} {
        \foreach \x in {0} {
            \fill[vcolor] (\x-\squarewidth,\y+2*\squarewidth) rectangle (\x+\squarewidth,\y-2*\squarewidth);
        }
    }
    
    \def\yshift{-.125}
    \foreach \x in {0.25} {
        \foreach \y in {.25,.5,.75,1,1.25} {
            \fill[hcolor] (\x-2*\squarewidth,\y+\yshift+\squarewidth) rectangle (\x+2*\squarewidth,\y+\yshift-\squarewidth);
        }
    }
    
    \fill[overlapcolor] (-\squarewidth,0.25+2*\squarewidth) rectangle (+\squarewidth,+0.25-2*\squarewidth);
    \fill[overlapcolor] (-\squarewidth,1+2*\squarewidth) rectangle (\squarewidth,+1-2*\squarewidth);

    \draw[\circlehighlight] (-1.5*\squarewidth,0.25+2.5*\squarewidth) rectangle (+1.5*\squarewidth,+0.25-2.5*\squarewidth);
    \draw[\circlehighlight] (.25-2.5*\squarewidth,0.5+\yshift+1.5*\squarewidth) rectangle (.25+2.5*\squarewidth,+0.5+\yshift-1.5*\squarewidth);
    \draw[\circlehighlight] (.25-2.5*\squarewidth,0.25+\yshift+1.5*\squarewidth) rectangle (.25+2.5*\squarewidth,+0.25+\yshift-1.5*\squarewidth);
\end{scope}

\def\arrowstart{0.5}
\draw[->] (\arrowstart,\arrowvertpos) -- (\arrowstart+\arrowlength,\arrowvertpos);

\begin{scope}[xshift=1cm]
  \clip (-.125,0) rectangle (0.35,1.35);

    \foreach \y in {.25,.5,.75,1,1.25} {
        \foreach \x in {0} {
            \fill[vcolor] (\x-\squarewidth,\y+2*\squarewidth) rectangle (\x+\squarewidth,\y-2*\squarewidth);
        }
    }
    
    \def\yshift{-.125}
    \foreach \x in {0.25} {
        \foreach \y in {.25,.5,.75,1,1.25} {
            \fill[hcolor] (\x-2*\squarewidth,\y+\yshift+\squarewidth) rectangle (\x+2*\squarewidth,\y+\yshift-\squarewidth);
        }
    }
    \fill[overlapcolor] (-\squarewidth,0.25+2*\squarewidth) rectangle (+\squarewidth,+0.25-2*\squarewidth);
    \fill[overlapcolor] (-\squarewidth,1+2*\squarewidth) rectangle (\squarewidth,+1-2*\squarewidth);

    \draw[\circlehighlight] (-1.5*\squarewidth,0.25+2.5*\squarewidth) rectangle (+1.5*\squarewidth,+0.25-2.5*\squarewidth);
    \draw[\circlehighlight] (.25-2.5*\squarewidth,0.5+\yshift+1.5*\squarewidth) rectangle (.25+2.5*\squarewidth,+0.5+\yshift-1.5*\squarewidth);
    \draw[\circlehighlight] (.25-2.5*\squarewidth,0.25+\yshift+1.5*\squarewidth) rectangle (.25+2.5*\squarewidth,+0.25+\yshift-1.5*\squarewidth);

    \foreach \y in {.25,.5,.75,1,1.25} {
        \draw[red, line width = \highlightwidth pt, opacity = \highlightopacity, line cap = round, scale = \scale] (0,\y) -- (0.25, \y+\yshift);
    }
\end{scope}

\def\arrowstart{1.5}
\draw[->] (\arrowstart,\arrowvertpos) -- (\arrowstart+\arrowlength,\arrowvertpos);

\begin{scope}[xshift=2cm]
  \clip (-.125,0) rectangle (0.35,1.35);

    \foreach \y in {.25,.5,.75,1,1.25} {
        \foreach \x in {0} {
            \fill[vcolor] (\x-\squarewidth,\y+2*\squarewidth) rectangle (\x+\squarewidth,\y-2*\squarewidth);
        }
    }
    
    \def\yshift{-.125}
    \foreach \x in {0.25} {
        \foreach \y in {.25,.5,.75,1,1.25} {
            \fill[hcolor] (\x-2*\squarewidth,\y+\yshift+\squarewidth) rectangle (\x+2*\squarewidth,\y+\yshift-\squarewidth);
        }
    }
    \fill[overlapcolor] (-\squarewidth,0.25+2*\squarewidth) rectangle (+\squarewidth,+0.25-2*\squarewidth);
    \fill[overlapcolor] (-\squarewidth,1+2*\squarewidth) rectangle (\squarewidth,+1-2*\squarewidth);

    \fill[overlapcolor] (+0.25-2*\squarewidth,0.25+\yshift+\squarewidth) rectangle (0.25+2*\squarewidth,\yshift+0.25-\squarewidth);
    \fill[overlapcolor] (+0.25-2*\squarewidth,\yshift+1+\squarewidth) rectangle (0.25+2*\squarewidth,\yshift+1-\squarewidth);

    \draw[\circlehighlight] (-1.5*\squarewidth,0.25+2.5*\squarewidth) rectangle (+1.5*\squarewidth,+0.25-2.5*\squarewidth);
    \draw[\circlehighlight] (.25-2.5*\squarewidth,0.5+\yshift+1.5*\squarewidth) rectangle (.25+2.5*\squarewidth,+0.5+\yshift-1.5*\squarewidth);
    \draw[\circlehighlight] (.25-2.5*\squarewidth,0.25+\yshift+1.5*\squarewidth) rectangle (.25+2.5*\squarewidth,+0.25+\yshift-1.5*\squarewidth);
    
    \foreach \y in {0,.25,.5,.75,1,1.25} {
        \draw[red, line width = \highlightwidth pt, opacity = \highlightopacity, line cap = round, scale = \scale] (0,\y) -- (0.25, \y+0.25+\yshift);
    }

\end{scope}

\def\arrowstart{2.5}
\draw[->] (\arrowstart,\arrowvertpos) -- (\arrowstart+\arrowlength,\arrowvertpos);

\begin{scope}[xshift=3cm]
  \clip (-.125,0) rectangle (0.35,1.35);

    \foreach \y in {.25,.5,.75,1,1.25} {
        \foreach \x in {0} {
            \fill[vcolor] (\x-\squarewidth,\y+2*\squarewidth) rectangle (\x+\squarewidth,\y-2*\squarewidth);
        }
    }
    
    \def\yshift{-.125}
    \foreach \x in {0.25} {
        \foreach \y in {.25,.5,.75,1,1.25} {
            \fill[hcolor] (\x-2*\squarewidth,\y+\yshift+\squarewidth) rectangle (\x+2*\squarewidth,\y+\yshift-\squarewidth);
        }
    }
    \fill[overlapcolor] (-\squarewidth,0.25+2*\squarewidth) rectangle (+\squarewidth,+0.25-2*\squarewidth);
    \fill[overlapcolor] (-\squarewidth,1+2*\squarewidth) rectangle (\squarewidth,+1-2*\squarewidth);

    \fill[overlapcolor] (+0.25-2*\squarewidth,0.25+\yshift+\squarewidth) rectangle (0.25+2*\squarewidth,\yshift+0.25-\squarewidth);
    \fill[overlapcolor] (+0.25-2*\squarewidth,\yshift+0.5+\squarewidth) rectangle (+0.25+2*\squarewidth,\yshift+0.5-\squarewidth);
    \fill[overlapcolor] (+0.25-2*\squarewidth,\yshift+1+\squarewidth) rectangle (0.25+2*\squarewidth,\yshift+1-\squarewidth);
    \fill[overlapcolor] (0.25-2*\squarewidth,\yshift+1.25+\squarewidth) rectangle (0.25+2*\squarewidth,\yshift+1.25-\squarewidth);

    \draw[\circlehighlight] (-1.5*\squarewidth,0.25+2.5*\squarewidth) rectangle (+1.5*\squarewidth,+0.25-2.5*\squarewidth);
    \draw[\circlehighlight] (.25-2.5*\squarewidth,0.5+\yshift+1.5*\squarewidth) rectangle (.25+2.5*\squarewidth,+0.5+\yshift-1.5*\squarewidth);
    \draw[\circlehighlight] (.25-2.5*\squarewidth,0.25+\yshift+1.5*\squarewidth) rectangle (.25+2.5*\squarewidth,+0.25+\yshift-1.5*\squarewidth);
    \draw[\circlehighlight] (0.25,1+\yshift) circle (\circlewidth pt);  
    \draw[\circlehighlight] (0.25,1.25+\yshift) circle (\circlewidth pt);
\end{scope}

\end{tikzpicture}
    \caption{An example vertical Dehn twist implemented in 2 rounds of $\operatorname{CNOT}$ gates. The layout is the same as in \cref{fig:toric layout} but with only vertical qubits (blue vertical rectangles) and horizontal qubits (green horizontal rectangles) from the column $(*,k),k\in C_m$ included for clarity. $\operatorname{CNOT}$ gates from vertical to horizontal qubits are represented by the red highlights. Boxes highlight the overlap of the Dehn twist with $\bar{\operatorname{X}}^h$ and the target logical~$\bar{\operatorname{X}}^v$ to be implemented conditioned on this overlap, defining the twist. Red vertices show how the Pauli-$\operatorname{X}$ propagates. A return to the original code is guaranteed by \cref{eqn: vanishing add term}. The twist overlaps on multiple horizontal logical operators, resulting in a simultaneous logical operation.}
    \label{fig:example dehn twist scheme}
\end{figure}

Perhaps the easiest way to show the generalised Dehn twist is through a visual example. Say we twist on some column indexed by $(*,k)$, that overlaps vertical edges $(e_l,k)$ on the logical $\bar{\operatorname{X}}^h_{1,j}$ and $(y^3,k)$ on the logical $\bar{\operatorname{X}}^h_{4,j}$. We want to implement a vertical logical defined as $\bar{\operatorname{X}}^v_{1,k}$ indexed by the horizontal edges $(e_l+y,k)$ conditioned on the overlapping edge $(e_l,k)$. A visual representation is shown in \cref{fig:example dehn twist scheme}.

It is clear from the figure that this twist maps
\begin{subequations}
    \begin{align}
        \bar{\operatorname{X}}^h_{1,j} &\mapsto\bar{\operatorname{X}}^h_{1,j}\bar{\operatorname{X}}^v_{1,j},\\
        \bar{\operatorname{X}}^h_{4,j} &\mapsto\bar{\operatorname{X}}^h_{4,j}\bar{\operatorname{X}}^v_{4,k}.
    \end{align}
\end{subequations}
An example of a different Dehn twist is to implement the logical $\bar{\operatorname{X}}^v_{4,k}$, circled in the final diagram of \cref{fig:example dehn twist scheme}, conditioned on the boxed vertical edge $(e_l,k)$. This maps
\begin{subequations}
    \begin{align}
        \bar{\operatorname{X}}^h_{1,j} &\mapsto \bar{\operatorname{X}}^h_{1,j}\bar{\operatorname{X}}^v_{4,k},\\
        \bar{\operatorname{X}}^h_{4,j} &\mapsto \bar{\operatorname{X}}^h_{4,j}\bar{\operatorname{X}}^v_{7,k},
    \end{align}
\end{subequations}
and can be implemented by applying $\operatorname{CNOT}$ gates from the boxed vertical edge to the circled horizontal edges sequentially. If the logical operator $\bar{\operatorname{X}}^v_{7,j}$ is not in the basis, it will have to be translated into the basis using the properties of the cyclic code.
\section{Dehn twists on $[[18q^2,8,2q]]$ Code} \label{sec:example}
We illustrate the generalised scheme with an in-depth example, for a hypergraph product code family that scales as $[[18q^2,8,2q]]$. We first describe the code construction and its properties. Then, we give an example of a Dehn twist and how it maps the logical operators. Lastly, we discuss the numerical simulation of the twist, and how the twist can be implemented fault-tolerantly.
\subsection{Description of the $[[18q^2,8,2q]]$ Code} \label{sec: tensor product 450 8 10 code}
\begin{figure}
    \centering
    \begin{tikzpicture}[scale=2, line width = 2pt]
        \colorlet{vertexcolor}{black} %
        \colorlet{linecolor}{orange!70!black}
        \colorlet{hcolor}{green!50!black} %
        \colorlet{vcolor}{blue!80!black}
        \colorlet{xvlog}{yellow!70!black}
        \colorlet{xhlog}{teal!80!black}
        \def\highlightwidth{10}
        \def\highlightopacity{0.5}
        \def\dashstart{-0.5}
        \def\dashend{-0.25}
        \def\linestart{-0.15}
        \def\lineend{0.15}
        \def\nodescale{1}
        \def\scale{1}
        \def\squarewidth{0.035}
        \def\trianglewidth{0.1}
        \def\root{1.73205080757}
        \foreach \x in {0,1,2} {
            \foreach \y in {0,1,2} {
                \fill[hcolor] (\x+0.5-2*\squarewidth,\y+\squarewidth) rectangle (\x+0.5+2*\squarewidth,\y-\squarewidth);
            }
        }

        \node[vcolor,anchor = north, scale = \nodescale] at (0,0+0.5) {$(e_l,e_l)$};
        \node[vcolor,anchor = north, scale = \nodescale] at (0,1+0.5) {$(x,e_l)$};
        \node[vcolor,anchor = north, scale = \nodescale] at (0,2+0.5) {$(x^2,e_l)$};

        \node[vcolor,anchor = north, scale = \nodescale] at (1,0+0.5) {$(e_l,x)$};
        \node[vcolor,anchor = north, scale = \nodescale] at (1,1+0.5) {$(x,x)$};
        \node[vcolor,anchor = north, scale = \nodescale] at (1,2+0.5) {$(x^2,x)$};

        \node[vcolor,anchor = north, scale = \nodescale] at (2,0+0.5) {$(e_l,x^2)$};
        \node[vcolor,anchor = north, scale = \nodescale] at (2,1+0.5) {$(x,x^2)$};
        \node[vcolor,anchor = north, scale = \nodescale] at (2,2+0.5) {$(x^2,x^2)$};

        \node[hcolor,anchor = west, scale = \nodescale] at (0+0.5,0) {$(e_l,e_l)$};
        \node[hcolor,anchor = west, scale = \nodescale] at (0+0.5,1) {$(x,e_l)$};
        \node[hcolor,anchor = west, scale = \nodescale] at (0+0.5,2) {$(x^2,e_l)$};

        \node[hcolor,anchor = west, scale = \nodescale] at (1+0.5,0) {$(e_l,x)$};
        \node[hcolor,anchor = west, scale = \nodescale] at (1+0.5,1) {$(x,x)$};
        \node[hcolor,anchor = west, scale = \nodescale] at (1+0.5,2) {$(x^2,x)$};

        \node[hcolor,anchor = west, scale = \nodescale] at (2+0.5,0) {$(e_l,x^2)$};
        \node[hcolor,anchor = west, scale = \nodescale] at (2+0.5,1) {$(x,x^2)$};
        \node[hcolor,anchor = west, scale = \nodescale] at (2+0.5,2) {$(x^2,x^2)$};

        \foreach \x in {0,1,2} {
            \foreach \y in {0,1,2} {
                \fill[vcolor] (\x-\squarewidth,\y+0.5+2*\squarewidth) rectangle (\x+\squarewidth,\y+0.5-2*\squarewidth);
            }
        }

        \foreach \x in {0,1} {
            \foreach \y in {0,1} {
            \draw[xhlog, line width = \highlightwidth pt, opacity = \highlightopacity, line cap = round, scale = \scale] (\x+\linestart,.5+\y) -- (\x+1+\lineend, .5+\y);
            }
        }
    
        \foreach \x in {0,1} {
            \foreach \y in {0,1} {
                \draw[xvlog, line width = \highlightwidth pt, opacity = \highlightopacity, line cap = round, scale = \scale] (\x+0.5,0+\y+\linestart) -- (\x+0.5, 1+\y+\lineend);
            }
        }
\end{tikzpicture}   
    \caption{Visual representation of $\bar{\operatorname{X}}_{i,j}^h$ and $\bar{\operatorname{X}}_{i,j}^v$ for the $[[18q^2,8,2q]]$ hypergraph product code family with $q=1$, and $1\leq i,j \leq2$. Yellow and teal highlights correspond to vertical and horizontal logical operators respectively. Each highlighted set represents a logical operator, and there are overlapping highlights on horizontal edges $\{(x,e_l),(x,x)\}$, and vertical edges $\{(e_l,x),(x,x)\}$. Dehn twists on rows or columns with overlapping logical operators act on the logical operators simultaneously. Codes with different $q$ extend this patch straightforwardly, with logical operators that overlap on the same edges on columns $(*,e_l)$ and $(*,x)$ and rows $(e_l,*)$ and $(x,*)$. Therefore the same set of Dehn twists apply to all codes in this family.}
    \label{fig:1+x+x^2 code logicals layout}
\end{figure}
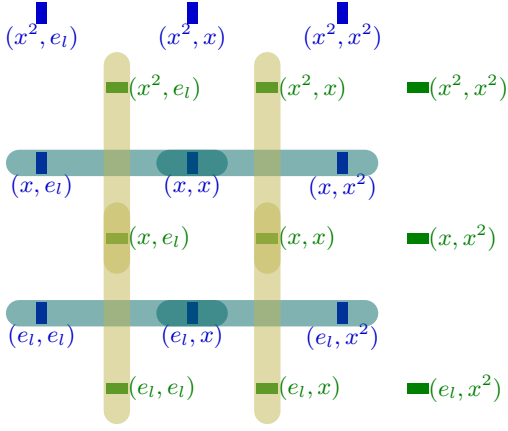

This code can be thought of as the simplest extension of the toric code, as it uses the check polynomial
\begin{equation}
    p(x) = e_l+x+x^2
\end{equation}
This polynomial is an irreducible polynomial for cyclic groups of order $3q,q\in\mathbb{N}$, and is the simplest polynomial with higher degree than that for the repetition code. The polynomial describes a general BCH code \cite[Chapter 3]{bok:MW}, a type of cyclic code, which guarantees a classical code distance of $2q$ \cite[Ch.~7.\ $\S$~6.\ Theorem 8]{bok:MW}.
The code is defined by the product of two cyclic groups $C_l=C_m = \langle x \rangle$. Let $l=3q$ for integer $q$. Then the check and generator polynomials are 
\begin{subequations} \label{eqn: check gen pols 90 8 10}
    \begin{align}
        p(x) &= e_l+x+x^2,\\
        g(x) &= (e_l+x)\sum_{i=1}^{q}x^{3i}.
    \end{align}
\end{subequations}
Like in the toric code, $p^T(x)$ and $p(x)$ describe the same classical code. From \cref{eqn: general logical basis} we readily write down the logical basis, where $1\leq i,j \leq 2$. For this example, we also use $g(x)$ for both the $\bar{\operatorname{X}}$ and $\bar{\operatorname{Z}}$ bases instead of $g^T(x)$ for one of them. There are 4 horizontal and vertical operators, for a total of 8 logical qubits. An illustration for the set of $\operatorname{\bar{X}}$ is given for $q=1$ in \cref{fig:1+x+x^2 code logicals layout}.

We consider the subset of twists to and from columns that fully support both $\bar{\operatorname{X}}$ and $\bar{\operatorname{Z}}$ logical operators. There are $\alpha_p$ such columns we can twist from, and $\alpha_p$ columns to twist to, indexed by the tuples $(*,e_l)$ and $(*,x)$. For each combination, there are $\alpha_p$ $\operatorname{CNOT}$ vertical Dehn twists that can be implemented; and finally, there are 2 types of twists: vertical and horizontal. Hence, there are $2\alpha_p^3=16$ twists in this subset.

The form of the generator polynomial for this hypergraph product  code family given by \cref{eqn: check gen pols 90 8 10} mean that all codes from the family have logical operators that overlap the same way on the rows and columns labelled by $e_l$ and $x$. This is depicted in \cref{fig:1+x+x^2 code logicals layout}. Therefore this set of Dehn twists apply to all codes in this code family.

A numerical simulation of the twists show that they generate the $\operatorname{PSL}(8,2)$ group on the logical qubit space. 
Here we point out again that the qubit layout is that of a torus due to the classical codes coming from a cyclic group algebra.
However, unlike the toric code, there is a larger set of generalised Dehn twists than generate a larger group of logical operations beyond the $\operatorname{Sp}(2,2)$ group. Effectively, by using different cyclic codes to construct the quantum code, the handle now supports more logical qubits and hence more Dehn twists.
We can also compare the code with a topological code derived from surface with $g=4$ handles, which also supports 8 logical qubits. On this code, Dehn twists would generate the $\operatorname{Sp}(8,2)$ group on the logical qubit space, which is a smaller set of logical operations.
\subsection{Example Dehn Twist on the Code}
Consider a twist from column $(*,x)$ to $(*,e_l)$, implementing the vertical logical $\bar{\operatorname{X}}^v_{2,1}$ conditioned on overlap on the vertical edge $(e_l, x)$. The logical operators $\bar{\operatorname{X}}^h_{1,1},\bar{\operatorname{X}}^h_{1,2}$ overlap with the twist on this edge, so by definition this provides the mapping
\begin{subequations}
    \begin{align}
        \bar{\operatorname{X}}^h_{1,1} &\mapsto \bar{\operatorname{X}}^h_{1,1}\bar{\operatorname{X}}^v_{2,1}\\
        \bar{\operatorname{X}}^h_{1,2} &\mapsto \bar{\operatorname{X}}^h_{1,2}\bar{\operatorname{X}}^v_{2,1},
    \end{align}
\end{subequations}
ignoring the identity maps on other logical operators.\\
\\Now consider the vertically-translated logical operators $\bar{\operatorname{X}}^h_{2,1}, \bar{\operatorname{X}}^h_{2,2}$, which overlap with the twist on the vertical edge $(x,x)$. The twist propagates the overlapping Pauli $\operatorname{X}$ to $\bar{\operatorname{X}}^v_{3,1}$, where there is an additional vertical translation due to the $x$-vertical translation of the overlapping edge.\\
\\The logical $\bar{\operatorname{X}}^v_{3,1}$ is not in our given basis. However, by property of this cyclic code,
\begin{equation}
    x^2 g(x) = (e_l+x) g(x).
\end{equation}
Therefore we can translate the implemented logical down,
\begin{equation}
    \bar{\operatorname{X}}^v_{3,1} = \bar{\operatorname{X}}^v_{1,1}\bar{\operatorname{X}}^v_{2,1}.
\end{equation}
The full mapping on $\bar{\operatorname{X}}$ for this twist is given as
\begin{subequations}\label{eqn:  example twist x mapping}
    \begin{align}
        \bar{\operatorname{X}}^h_{1,1} &\mapsto \bar{\operatorname{X}}^h_{1,1}\bar{\operatorname{X}}^v_{2,1},\\
        \bar{\operatorname{X}}^h_{1,2} &\mapsto \bar{\operatorname{X}}^h_{1,2}\bar{\operatorname{X}}^v_{2,1},\\
        \bar{\operatorname{X}}^h_{2,1} &\mapsto \bar{\operatorname{X}}^h_{2,1}\bar{\operatorname{X}}^v_{2,1}\bar{\operatorname{X}}^v_{1,1},\\
        \bar{\operatorname{X}}^h_{2,2} &\mapsto \bar{\operatorname{X}}^h_{2,2}\bar{\operatorname{X}}^v_{2,1}\bar{\operatorname{X}}^v_{1,1}.
    \end{align}
\end{subequations}
More generally, the effect of Dehn twists can be obtained by considering how the shifted polynomial $x^kg(x)$ for some $k$ is decomposed in the given basis of $\mathcal{H}_1$ for the cyclic code.\\
\\In terms of the action on the coboundary and boundary operators, note that when target and control columns are different, the additional term in the coboundary operator no longer has the identity element in the tensor product. Instead, it is replaced by $x^{i-j}$ if we twist from column $x^j$ to $x^i$. The general form of the coboundary operator in the $k^{th}$ intermediate step is given by 
\begin{equation} \label{eqn: general twist cob operator}
\renewcommand\arraystretch{1.2}
    \delta_0^{(k)}|_{twist} =\left[ 
        \begin{array}{c}
            p_1^T(x)\otimes e_l\\\hline
             e_l\otimes p_2^T(x) + p^T_1(x)g^{(k)}(x)\otimes x^{i-j},
        \end{array}
        \right].
\end{equation}

In this example, the first step of the Dehn twist gives the local coboundary operator
\begin{equation} \label{eqn: example twist cob operator2}
\renewcommand\arraystretch{1.2}
    \delta_0^{(1)}|_{twist} =\left[ 
        \begin{array}{c}
            p_1^T(x)\otimes e_l\\\hline
             e_l\otimes p_2^T(x) + p^T_1(x)(x)\otimes x^{-1}
        \end{array}
        \right].
\end{equation}

We can consider the effect on the $\bar{\operatorname{Z}}$- logical operators by swapping the control and target qubit set for the $\operatorname{CNOT}$, and noting that implementing $g(x)$ for $\bar{\operatorname{X}}$ implements $g^T(x)$ for $\bar{\operatorname{Z}}$ by symmetry. In this example,
\begin{equation}
    g^T(x) = (e_l+x)g(x),
\end{equation}
and factoring in the vertical translations from the overlaps of the twist column and the logical operators gives us the mapping
\begin{subequations} \label{eqn:  example twist z mapping}
    \begin{align}
        \bar{\operatorname{Z}}^h_{1,1} &\mapsto \bar{\operatorname{Z}}^h_{1,1}\bar{\operatorname{Z}}^v_{2,2},\\
        \bar{\operatorname{Z}}^h_{2,1} &\mapsto \bar{\operatorname{Z}}^h_{2,1}\bar{\operatorname{Z}}^v_{2,2}\bar{\operatorname{Z}}^v_{1,2}.
    \end{align}
\end{subequations}
Due to how the logical basis is defined, the effect of horizontal twists is easily obtained by swapping superscripts $h$ and $v$, and subscripts $i,j$ in \cref{eqn:  example twist x mapping,eqn:  example twist z mapping}. Explicitly the horizontal twist from row $(x,*)$ to $(e_l,*)$ gives the map
\begin{subequations}\label{eqn:  example twist all mapping horizontal}
    \begin{align}
        \bar{\operatorname{X}}^v_{1,1} &\mapsto \bar{\operatorname{X}}^v_{1,1}\bar{\operatorname{X}}^h_{1,2},\\
        \bar{\operatorname{X}}^v_{2,1} &\mapsto \bar{\operatorname{X}}^v_{2,1}\bar{\operatorname{X}}^h_{1,2},\\
        \bar{\operatorname{X}}^v_{1,2} &\mapsto \bar{\operatorname{X}}^v_{1,2}\bar{\operatorname{X}}^h_{1,2}\bar{\operatorname{X}}^h_{1,1},\\
        \bar{\operatorname{X}}^v_{2,2} &\mapsto \bar{\operatorname{X}}^v_{2,2}\bar{\operatorname{X}}^h_{1,2}\bar{\operatorname{X}}^h_{1,1},\\
        \bar{\operatorname{Z}}^v_{1,1} &\mapsto \bar{\operatorname{Z}}^v_{1,1}\bar{\operatorname{Z}}^h_{2,2},\\
        \bar{\operatorname{Z}}^v_{1,2} &\mapsto \bar{\operatorname{Z}}^v_{1,2}\bar{\operatorname{Z}}^h_{2,1}\bar{\operatorname{Z}}^h_{2,2}.
    \end{align}
\end{subequations}
We obtain the full set of 16 twists from similar arguments. 
\subsection{Numerical Implementation of the Dehn Twists}
The twists were numerically simulated for integers $3\leq q \leq 6$, corresponding to distance $d\in\{6,8,10,12\}$ hypergraph product codes. A basis change was also applied to the $\bar{\operatorname{X}}$ logical operators in \cref{eqn: general logical basis} while keeping $\bar{\operatorname{Z}}$ logical operators as written to obtain a symplectic basis.

While the underlying polynomial is always ${e_l+x+x^2}$, choosing different representative polynomials can lead to different intermediate code distances, which can be reduced compared to the original code. The polynomials also affect the maximum intermediate check weight.

Selecting $p(x)=e_l+x+x^2$ for both $p_1(x)$ and $p_2(x)$ results in intermediate codes that maintain the original code distances. We also found that the maximum check weight during the operation was~9 for all codes; and this maximum checkweight is valid for simulations up to $q\leq200$. This is due to the repetitive nature of the generator polynomial given in \cref{eqn: check gen pols 90 8 10}. %
Moreover, for the codes that we tested, the twists together with fold-transversal gates \cite{breuckmann2022foldtransversal} generate the full Clifford group on the logical space.
\section{Balanced Product Cyclic Codes} \label{sec: balanced product}
In this section, we extend the general Dehn twist scheme towards the \textit{balanced} product of cyclic codes. We first discuss how the Dehn twist for balanced product cyclic codes differ from that of the hypergraph product, and how they can be applied.
We then give an example code family, constructed by taking the balanced product of two copies of cyclic groups $C_{3q}$ over the cyclic subgroup $C_q$. This is similar to the construction for the $[[18q^2,8,2q]]$ code detailed in \cref{sec: tensor product 450 8 10 code}, but with parameters $[[n,k]] = [[18q,8]]$, and with distance \textit{upper-bounded} by $2q$. 
For this code family we present results of a numerical search for codes with check weight 6 and 8. We investigate the $[[90,8,10]]$ code found by taking $q=5$ and factoring out a $C_5$ cyclic subgroup under the balanced product, and use it to show that the Dehn twists considered for the hypergraph product apply straightforwardly to this family of balanced product codes for odd $q$. Finally we show that the $[[90,8,10]]$ bivariate bicycle code from \cite{bravyi2023highthreshold} is from this code family. In doing so, we not only show that the generalised Dehn twists also apply to the bivariate bicycle code, but also provide a logical basis as described in \cref{eqn: general logical basis} for their code, which was only found previously through numerical simulation.
\subsection{Extension of Dehn Twists to Balanced Products}
The generalised Dehn twist scheme extends naturally to the balanced product of cyclic codes. The code is obtained by taking a balanced product of the group algebras over some subgroup $H$ instead of over $\mathbb{F}_2$. In linear-algebraic terms, a balanced product of two vector spaces $V$ and $W$ equipped with a linear right (resp. left) action of $H$ is the quotient \cite{Breuckmann_2021_2}
\begin{equation}
    V\otimes_HW = V\otimes W/\langle vh\otimes w - v\otimes hw\rangle
\end{equation}
for $v\in V, w\in W$. In this case, $V$ and $W$ are $\mathbb{F}_2C_l$ and~$\mathbb{F}_2C_m$. The balanced product partitions elements of the vector space into equivalence classes, which are orbits of $H$ with an anti-diagonal action. Effectively, this replaces every group algebra element $a\otimes b$ in previous sections with the element $a\otimes_Hb$, where
\begin{equation}
    a\otimes_Hb = \{ah\otimes h^{-1}b\mid h\in H\}.
\end{equation}

The logical bases and Dehn twists remain valid because \cref{eqn: vanishing add term} still holds when considering equivalence classes of the group algebra elements under the balanced product. 
Therefore, the Dehn twists considered for the hypergraph product codes are shown to apply straightforwardly.

Furthermore, due to the K\"{u}nneth formula for balanced products \cite[Theorem 19]{Breuckmann_2021_2} the logical basis in \cref{eqn: general logical basis} is still valid, albeit with the tensor products~$\otimes$ replaced by balanced products~$\otimes_H$. 
Under the balanced product, some logical operators might be mapped to the same equivalence class, therefore reducing $\dim\mathcal{H}_1$ of the quantum code; and the logical basis given in \cref{eqn: general logical basis} might not be minimum weight, so the choice of representative polynomials for the input classical codes matter. 
Also, in general, $\operatorname{CNOT}$ operations are no longer confined to single columns or rows. This might result in 0 overlapping edges modulo 2 between a twist column or row and the logical operators. 
Should this happen, the Dehn twist returns us to the same quantum code, but applies an identity mapping on the logical operators.

Therefore, to determine the effect of Dehn twists, in general one must consider how the rows or columns translate under the balanced product, and then determine the overlap modulo 2 with a basis of logical operators.
\subsection{Constructing a $[[18q,8,\leq 2q]]$ Balanced Product Cyclic Code}
We can construct this family of codes using the check and generator polynomials as described in \cref{eqn: check gen pols 90 8 10} for the $[[18q^2,8,2q]]$ hypergraph product code, using two copies of a $C_{3q}=\langle x\rangle$ cyclic group. However, instead of taking a tensor product over $\mathbb{F}_2$, we factor out a $C_q=\langle x^3 \rangle$ cyclic subgroup under the balanced product. The hypergraph product code distance of $2q$ upper-bounds the distance of this construction as the basis of logical operators from \cref{eqn: general logical basis} is still valid. However, the minimum-weight logical operators of the balanced product code now depend on representative polynomials $p_1(x), p_2(x)$ used in the construction.

The effect of factoring out $C_q$ is to divide the total qubit number $18q^2$ by $q$ as the subgroup acts freely and transitively on $C_{3q}$. Therefore every element in the tensor product group algebra is partitioned into an equivalence class, and there are no fixed points.

The dimensions of the homology and cohomology groups of the total complex, and therefore the quantum code, do not change. We see this from the basis of logical operators given in \cref{eqn: general logical basis}. Another way of stating the effect of partitioning group algebra elements into equivalence classes is that it allows movement of elements in $\langle x^3\rangle$ across the tensor product. This movement might make some logical operators equivalent. In our example it does not happen as $\deg p(x)=2<3$, the degree of the subgroup generator. Therefore the homology group representatives given by \cref{eqn: homology group representatives} under the balanced product remain as linearly-independent vectors in distinct equivalence classes.

When $3q$ is odd, the subgroup $\langle x^3 \rangle$ is in fact the largest subgroup that can be factored out under the balanced product without reducing homology group dimensions, as the monomial $x^2$ generates the full group. In general this has a non-trivial effect on the balanced product code.\footnote{For example, the balanced product code from two $[7,4,3]$ Hamming codes from the $\mathbb{F}_2C_7$ group algebra, factoring out $C_7$ has two sets of parameters. 
If $p_1(x)=p_2(x)$, the returned code has parameters $[[14,8,2]]$. If $p_1(x)\neq p_2(x)$, it is $[[14,2,3]]$, effectively two copies of a Steane code.}
We leave a detailed analysis for future work.

We can construct matrix representations of polynomials from the balanced product by defining a vector space, and identifying representative elements from each equivalence class to be the basis. For $H=C_q$ and $l=3q$, we consider the vector space
\begin{equation} \label{eqn:vecspace}
\begin{split}
    C_{l}\otimes_{H}C_{l}=\langle &e_l\otimes_He_l,\cdots,e_l\otimes_Hx^{l-1},\\
    &x\otimes_He_l,\cdots,x\otimes_Hx^{l-1},\\
    &x^2\otimes_He_l,\cdots,x^2\otimes_Hx^{l-1} \rangle.
\end{split}
\end{equation}

Under this basis, the generators of the group algebra can be mapped to block matrices
\begin{subequations} \label{eqn: bp group algebra generators}
    \begin{align}
        e_l\otimes_H x &\mapsto \begin{bmatrix}
            x & 0 & 0\\
            0 & x & 0\\
            0 & 0 & x
        \end{bmatrix}\\
        x \otimes_H e_l &\mapsto \begin{bmatrix} \label{eqn: bp group algebra generators xe}
            0 & 0 & x^3\\
            e_l & 0 & 0\\
            0 & e_l & 0
        \end{bmatrix}
    \end{align}
\end{subequations}

The $x^3$ entry in \cref{eqn: bp group algebra generators xe} is due to the fact that for $h\in C_{l}$,
\begin{equation}
    (x\otimes_He_l)(x^2\otimes_Hh) = e_l\otimes_Hx^3h.
\end{equation}
We simulate $l$ up to 100, and find that the generated group algebra is isomorphic,
\begin{equation} \label{eqn: isomorphism of bp to 3x3q group algebra}
    \mathbb{F}_2C_l\otimes_H\mathbb{F}_2C_l \cong \mathbb{F}_2C_3\otimes\mathbb{F}_2C_l.
\end{equation}
For $q\mod 3 \neq0$, there is a simple isomorphism: letting $C_3=\langle y \rangle,$ we can define
\begin{subequations} \label{isomorphism of balanced product to cyclic group algebra}
    \begin{align}
        f: \mathbb{F}_2 C_{l} \otimes_H \mathbb{F}_2 C_{l} &\to \mathbb{F}_2 {C_{3}}\times C_{l}\\
        f(x^q\otimes_H e_l) &\mapsto y\otimes e_l\\
        f(e_l\otimes_H x) &\mapsto e_3 \otimes x.
    \end{align}
\end{subequations}

A numerical search was conducted to find representative polynomials $p_1(x), p_2(x)$ with weight 3 that maximise the distance of the quantum code. These give a code with check weight 6. The first element of each polynomial is fixed to be $e_l$. The distance of the codes are upper-bounded by $2q$, which is the distance of the hypergraph product code. A summary of results is given in \cref{table: search results}. 
\begin{table}
\centering
\begin{tabularx}{\linewidth}{c|X|c|X|X|X}
 $q$ & $[[n,k,d]]$  & $2q$ & $p_1(x)$ & $p_2(x)$ \\\hline
\rowcolor{yellow}1 & [[18,8,2]] & 2 & $e_l+x+x^2$ & $e_l+x+x^2$\\
\rowcolor{yellow}2 & [[36,8,4]] & 4 & $e_l+x+x^2$ & $e_l+x+x^2$\\
3 & [[54,8,4]]& 6 & $e_l+x+x^2$ & $e_l+x+x^2$\\
\rowcolor{yellow}4 & [[72,8,8]]& 8 & $e_l+x+x^5$ & $e_l+x+x^8$ \\
\rowcolor{yellow}5 & [[90,8,10]]& 10 &$e_l+x+x^5$ & $e_l+x^2+x^7$\\
6 & [[108,8,8]] & 12 & $e_l+x+x^5$ & $e_l+x+x^2$ \\
7 & [[126,8,10]] & 14 & $e_l+x+x^5$ & $e_l+x^2+x^7$\\
8 & [[144,8,12]] & 16 & $e_l+x+x^5$ & $e_l+x+x^{11}$\\
9 & [[162,8,12]] & 18 & $e_l+x+x^5$ & $e_l+x+x^{11}$\\
10 & [[180,8,16]] & 20 & $e_l+x+x^5$ & $e_l+x+x^{17}$\\
\end{tabularx}
\caption{Highest-distance codes from a numerical search using weight 3 polynomials in the construction of ${[[18q,8,\leq2q]]}$ balanced product cyclic codes for $q\leq10$. Codes that saturate the upper bound of $2q$ are highlighted in yellow.}
\label{table: search results}
\end{table} 
We also attempted the construction using polynomials of higher weight. There are no weight 4 polynomials with $e_l+x+x^2$ as a common factor. 
Hence we try combinations of weight 3 and weight 5 polynomials, which give a quantum code of check weight 8. 
Results are shown in \cref{table: weight 3 and weight 5 polynomial numerical search}. Distances quoted are exact distances found by formulating distance-finding as a problem in linear integer programming, and solved by using \cite{gurobi}. The parity check matrices of all balanced product codes in \cref{table: search results,table: weight 3 and weight 5 polynomial numerical search} can be found at \cite{ryan_tiew_2024_14041286}.

\begin{table}
\centering
\begin{tabularx}{\linewidth}{c|X|c|X|c|c}
$q$ & $[[n,k,d]]$  & $2q$ & $p_1(x)$ & $p_2(x)$ \\\hline
\rowcolor{yellow} 3 & [[54,8,6]]& 6 & $e_l+x+x^2$ & $e_l+x+x^2+x^3+x^6$\\
\rowcolor{yellow} 6 & [[108,8,12]] & 12 & $e_l+x+x^5$ & $e_l+x+x^2+x^3+x^6$ \\
\rowcolor{yellow}7 & [[126,8,14]] & 14 & $e_l+x+x^8$ & $e_l+x+x^3+x^{17}+x^{18}$\\
\rowcolor{yellow}8 & [[144,8,16]] & 16 & $e_l+x+x^5$ & $e_l+x+x^3+x^8+x^{21}$\\
\end{tabularx}
\caption{Codes from the same construction as in \cref{table: search results} but using weight 3 and weight 5 polynomials, and $q\leq8$. The search was repeated for $q$ that did not saturate the upper bound from weight 3 polynomials alone. Increasing the polynomial weight, and hence the check weight of the quantum code, can increase distance. In this case, all codes with $q\leq8$ saturate the distance bound. Codes with higher~$q$ were not explored due to time and computational constraints. 
}
\label{table: weight 3 and weight 5 polynomial numerical search}
\end{table}
From the numerical search, we find some quantum codes that achieve maximum distance with check weight~6. In particular, we are able to construct a $[[90,8,10]]$ quantum code, which we describe in detail in the next section. We also show that the 16 Dehn twists as described in \cref{sec:example} also apply to these codes for \textit{odd} $q$, as the Dehn twist rows and columns will have an 1 overlapping edge modulo 2 with the logical operators.  
The tensor product of cyclic group algebras $\mathbb{F}_2C_l\otimes\mathbb{F}_2C_m$ corresponded to a $l\times m$ toric layout. The balanced product here describes the topology of a \textit{twisted} torus of height 3 and length $l$.
We also find that increasing the check weight by using higher-weight polynomials lets us find codes that saturate the distance upper bound $2q$, that could not be found using only weight 3 polynomials. We leave the search using polynomials of higher weight and for larger $q$ as future work.

\subsection{Dehn Twists on the $[[90,8,10]]$ Balanced Product Code} \label{sec: BP code example}
\begin{figure*}
    \centering
    \begin{tikzpicture}[scale=1.15, line width = 2pt,framed,background rectangle]
        \colorlet{vertexcolor}{black} %
        \colorlet{linecolor}{orange!70!black}
        \colorlet{hcolor}{green!50!black} %
        \colorlet{vcolor}{blue!80!black}
        \colorlet{xvlog}{yellow!80!black}
        \colorlet{xhlog}{teal!80!black}
        \def\highlightwidth{5}
        \def\highlightopacity{0.5}
        \def\dashstart{-0.5}
        \def\dashend{-0.25}
        \def\linestart{-0.25}
        \def\lineend{2.7}
        \def\nodescale{0.85}
        \def\scale{1}
        \def\squarewidth{0.035}
        \def\trianglewidth{0.1}
        \def\root{1.73205080757}
        \foreach \x in {0,1,2,3,4,5,6,7,8,9,10,11,12,13,14} {
            \foreach \y in {0,1,2} {
                \fill[hcolor] (\x+0.5-2*\squarewidth,\y+\squarewidth) rectangle (\x+0.5+2*\squarewidth,\y-\squarewidth);
            }
        }

        \node[vcolor,anchor = north, scale = \nodescale] at (0,0+0.5) {$(e_l,e_l)$};
        \node[vcolor,anchor = north, scale = \nodescale] at (0,1+0.5) {$(x,e_l)$};
        \node[vcolor,anchor = north, scale = \nodescale] at (0,2+0.5) {$(x^2,e_l)$};
        
        \node[vcolor,anchor = north, scale = \nodescale] at (3,0+0.5) {$(x^3,e_l)$};
        \node[vcolor,anchor = north, scale = \nodescale] at (3,1+0.5) {$(x^4,e_l)$};
        \node[vcolor,anchor = north, scale = \nodescale] at (3,2+0.5) {$(x^5,e_l)$};
    
        \node[vcolor,anchor = north, scale = \nodescale] at (6,0+0.5) {$(x^6,e_l)$};
        \node[vcolor,anchor = north, scale = \nodescale] at (6,1+0.5) {$(x^7,e_l)$};
        \node[vcolor,anchor = north, scale = \nodescale] at (6,2+0.5) {$(x^8,e_l)$};
        
        \node[vcolor,anchor = north, scale = \nodescale] at (9,0+0.5) {$(x^9,e_l)$};
        \node[vcolor,anchor = north, scale = \nodescale] at (9,1+0.5) {$(x^{10},e_l)$};
        \node[vcolor,anchor = north, scale = \nodescale] at (9,2+0.5) {$(x^{11},e_l)$};
    
        \node[vcolor,anchor = north, scale = \nodescale] at (12,0+0.5) {$(x^{12},e_l)$};
        \node[vcolor,anchor = north, scale = \nodescale] at (12,1+0.5) {$(x^{13},e_l)$};
        \node[vcolor,anchor = north, scale = \nodescale] at (12,2+0.5) {$(x^{14},e_l)$};
    
        \foreach \x in {0,1,2,3,4,5,6,7,8,9,10,11,12,13,14} {
            \foreach \y in {0,1,2} {
                \fill[vcolor] (\x-\squarewidth,\y+0.5+2*\squarewidth) rectangle (\x+\squarewidth,\y+0.5-2*\squarewidth);
            }
        }
    
        \def\patchsize{2}
        \draw[red, dashed, thick] (-.45,-.45) rectangle (.6+\patchsize,.6+\patchsize);
    
        \foreach \x in {0,3,6,9,12} {
            \draw[xhlog, line width = \highlightwidth pt, opacity = \highlightopacity, line cap = round, scale = \scale] (\x,.5) -- (\x+1, .5);
        }
    
        \foreach \x in {0,3,6,9,12} {
            \draw[xvlog, line width = \highlightwidth pt, opacity = \highlightopacity, line cap = round, scale = \scale] (\x+0.5,0) -- (\x+0.5, 1);
        }
\end{tikzpicture}   
    \caption{Qubits of a $[[90,8,10]]$ balanced product code laid out on a two-dimensional grid. The layout is the same as \cref{fig:toric layout} but contains only the subset of qubits as given in \cref{eqn:vecspace}. Qubits of the horizontal logical $\bar{\operatorname{X}}^h_{1,1}$ are highlighted in teal, and $\bar{\operatorname{X}}^v_{1,1}$ are highlighted in yellow. The rest of the logical operators can be obtained by shifting the highlights a single step to the right and up as in \cref{fig:1+x+x^2 code logicals layout}, for a total of 8 logical operators. The qubit layout is that of a twisted torus. The set of labelled vertical edges are defined by \cref{eqn: vertical column on c15 code} and represent a column of vertical edges $(*,e_m)$ for Dehn twists similar to the hypergraph product codes. However, due to the twisted vertical boundary conditions, the column now translates horizontally, equivalent to twisting along the torus when moving along the horizontal direction. The dashed red box indicates a patch of the torus that twists in this way. Under the balanced product, the logical $\bar{\operatorname{X}}^v_{1,1}$ now has horizontal translations. Finally, note that the $\bar{\operatorname{X}}$-logical for the $[[90,8,10]]$ bivariate bicycle code of \cite[Table 7]{bravyi2023highthreshold} is exactly the logical corresponding to $\bar{\operatorname{X}}^h_{1,1}\bar{\operatorname{X}}^h_{1,2}$ in our code.}
    \label{fig:c15_code_layout}
\end{figure*}
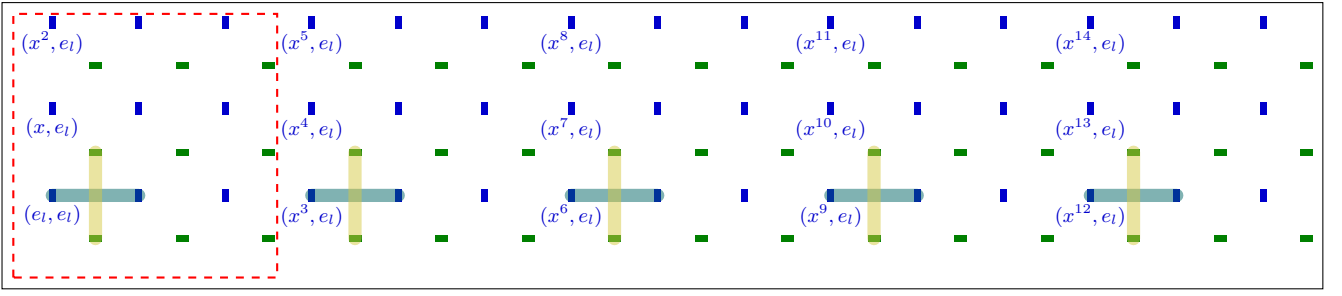

To illustrate Dehn twists on this class of balanced product cyclic codes, we look at the $[[90,8,10]]$ code from \cref{table: search results} where $q=5, l=15$. The code is defined by the representative polynomials
\begin{subequations} \label{eqn: BP representative polynomials}
    \begin{align}
        \begin{split}
            p_1(x) &= e_l+x+x^5 \\
            &= (e_l+x+x^2)(e_l+x^2+x^3),
        \end{split}\\
        \begin{split}
            p_2(x) &= e_l+x^2+x^7\\
            &= (e_l+x+x^2)(e_l+x^2+x^4+x^5),
        \end{split}
    \end{align}
\end{subequations}
and have check and generator polynomials as in \cref{eqn: check gen pols 90 8 10}, and a logical basis given by \cref{eqn: general logical basis}. Explicitly, in the basis of \cref{eqn:vecspace}, these polynomials have the block matrix representation
\begin{subequations}
    \begin{align}
        p_1(x)\otimes_He_l &\mapsto \begin{bmatrix}
            e_l & x^6 & x^3\\
            e_l & e_l & x^6\\
            x^3 & e_l & e_l
        \end{bmatrix},\\
        e_l\otimes_Hp_2(x) &\mapsto \begin{bmatrix}
            p_2(x) & 0 & 0\\
            0 & p_2(x) & 0\\
            0 & 0 & p_2(x)
        \end{bmatrix}.
    \end{align}
\end{subequations}
Constructing the boundary operators as in \cref{eqn:gen_operators} returns a $[[90,8,10]]$ quantum code with an explicit logical operator basis given by \cref{eqn: general logical basis}. A layout of this code is given in \cref{fig:c15_code_layout}, which we will refer to for the remainder of this section.

Consider a Dehn twist to implement $\bar{\operatorname{X}}^v$ conditioned on $\bar{\operatorname{X}}^h$. In the hypergraph product code, to implement a vertical twist, we considered $\operatorname{CNOT}$ gates from a column of vertical edges  $(*,h),h\in C_m$. In the $[[450,8,10]]$ code and taking $h=e_m$ for example, these edges are explicitly indexed as $\{e_l\otimes e_m, \cdots, x^{14}\otimes e_m\}$. 

Considering the twist for the balanced product code, the same column of vertical edges are identified as
\begin{equation} \label{eqn: vertical column on c15 code}
    \begin{split}
        \{e_l\otimes_H e_l, x\otimes_H e_l, x^2\otimes_H e_l,x^3\otimes_H e \cong e\otimes_H x^3,\\
        \cdots, e\otimes_H x^{12}, x\otimes_H x^{12}, x^2\otimes_H x^{12}\}.
    \end{split}
\end{equation}
Parts of the vertical column have now been shifted horizontally due to the partitioning of equivalence classes. This same shift also explains horizontal translations in the logical operators $\bar{\operatorname{X}}^v, \bar{\operatorname{Z}}^v$. Treating the column as a straight vertical line on the surface, we also see visually that factoring out the cyclic subgroup has changed the surface topology to that of a \textit{twisted} torus; this topology is valid for other balanced products of cyclic codes, factoring out a cyclic subgroup. A basis of logical operators follow naturally from \cref{eqn: general logical basis}, and by considering equivalence class representatives in our prescribed domain. Having identified the column, the Dehn twist scheme proceeds as described for the hypergraph product codes.

Compared to the hypergraph product code family, there are some differences. Firstly, in the hypergraph product code, the (vertical) Dehn twist is from single columns to single columns. In the balanced product code, by factoring out $\langle x^3\rangle$, edges in the column are translated horizontally, spreading them out on the toric layout. This leads to the second difference: there are a greater number of overlapping edges between the Dehn twists and the horizontal logical. In the hypergraph product code, there was a maximum of 1; in this example, there are 5, shown in \cref{fig:c15_code_layout}. However, because the number of overlapping edges is still 1 modulo 2, the implementation of the Dehn twists are the same as for the hypergraph product code. Generally, the effect of factoring out the subgroup on the number of overlapping edges needs to be considered to determine if the twist propagates logical operators in the same way as for the hypergraph product code.

In this case, we see from the twisted torus topology that the number of overlapping edges with the Dehn twist is exactly $q$. 
Increasing $q$ increases the length of the twisted torus, and the number of times the vertical logical operators and the Dehn twist column wraps around the torus. 
Effectively, it elongates the torus horizontally by the patch in \cref{fig:c15_code_layout} boxed in the red dashed lines. Therefore, while the set of 16 Dehn twists are valid for all $q$ in the $[[18q^2,8,2q]]$ hypergraph product family, it is only valid for odd $q$ in this balanced product cyclic code family.

Numerical simulations implementing the Dehn twist on this code show that intermediate code distances reduce to a minimum of~6, while the maximum check weight increases to a maximum of~16. The maximum check weight is dependent on the additional polynomial term in the boundary and coboundary operators. For example, in the horizontal twist scheme implementing $g(x)$, the coboundary operator 
\begin{equation} \label{eqn: example twist cob operator}
\renewcommand\arraystretch{1.2}
    \delta_0^{(6)}|_{twist} =\left[ 
    \begin{array}{c}
        p_1^T(x)\otimes e_l + e_l\otimes p_2^T(x)g^{(6)}(x)\\\hline
         e_l\otimes p_2^T(x)
    \end{array}
    \right]
\end{equation}
has the additional term
\begin{equation}
\begin{split}
    p_2^T(x)g^{(6)}(x) &= (e_l+x^{-2}+x^{-7})(e_l+x)(e_l+x^3+x^6)\\
    &= x^{13} + x^{12} + x^{11} + x^{9} + x^{8} \\
    &+ x^7 + x^6 + x^5 + x^3 + x^2.
\end{split}
\end{equation}
The additional term has weight 10. Adding the other coboundary terms returns weight~16 $\operatorname{X}$-checks restricted to the row.
In general the choice of representative polynomials is important for both determining the balanced product code distances, as well as intermediate code distances and check weights. Furthermore, we also find that, while the set of Dehn twists are still valid, the full logical Clifford group is no longer generated by the set of Dehn twists and fold-transversal gates. 
\subsection{$[[90,8,10]]$ Bivariate Bicycle Code as a Balanced Product}
We state here the $[[90,8,10]]$ bivariate bicycle code from \cite{bravyi2023highthreshold} and show it is in the family of balanced product cyclic codes considered in \cref{sec: BP code example}.

Bivariate bicycle codes are constructed from the group algebra $\mathbb{F}_2 C_{j}\otimes C_{k}$. To keep consistent with terminology used in this paper and prevent confusion between generators, let the cyclic groups be defined by
\begin{subequations}
    \begin{align}
        C_j &= \langle s \mid s^j=e_j\rangle,\\
        C_k &= \langle t \mid t^k=e_t\rangle.
    \end{align}
\end{subequations}
Then the tensor product group algebra has the generators
\begin{subequations}
\begin{align}
    \alpha &= s\otimes e_k,\\
    \beta &= e_j \otimes t.
\end{align}
\end{subequations}
For the bivariate bicycle $[[90,8,10]]$ code, let $j=15, {k = 3}$. The code is given by two polynomials,
\begin{subequations}
\begin{align}
    A &= \alpha^9 + \beta + \beta^2,\\
    B &= 1+ \alpha^2 + \alpha^7,
\end{align}
\end{subequations}
where 1 refers to the identity element $e_{15}\otimes e_3$. The parity check matrices are constructed as
\begin{subequations}
    \begin{align}
        H_X &= \left[ A \mid B \right],\\
        H_Z &= \left[ B^T \mid A^T\right].
    \end{align}
\end{subequations}
We can define an isomorphism via \cref{eqn: isomorphism of bp to 3x3q group algebra}, and with $q=5$. Explicitly, the map is given by 
\begin{subequations}
    \begin{align}
        f: \mathbb{F}_2 {C_{15}}\times C_3 &\to \mathbb{F}_2 C_{15} \otimes_H \mathbb{F}_2 C_{15}\\
        f(\alpha) &\mapsto x\otimes_H e_l,\\
        f(\beta) &\mapsto e_l \otimes_H x^5.
    \end{align}
\end{subequations}
This maps the bivariate bicycle code polynomials as
\begin{subequations}
    \begin{align}
        \begin{split}
            f(A) &\mapsto e_l\otimes_H (x^5+x^9+x^{10})\\
             &= e_l\otimes_H (e_l+x+x^2)(x^5)(e_l+x+x^3),
        \end{split}\\
        \begin{split}
            f(B) &\mapsto (e_l+x^2+x^7) \otimes_H e_l \\
            &= (e_l+x+x^2)(e_l+x+x^2+x^4+x^5) \otimes_H e_l.
        \end{split}
    \end{align}
\end{subequations}
The isomorphism shows that the logical basis for the balanced product code, which we defined analytically, apply to the bivariate bicycle code, whose logical basis was found numerically. The set of Dehn twists also apply to this code. However, with the form of the boundary operators defined in this paper, the $A$ and $B$ polynomials must be swapped around. This is simply a different way of stacking the vertical and horizontal boundary operators from the double complex to construct the boundary operators for the total complex.

We also numerically simulate the Dehn twists for the bivariate bicycle codes under this isomorphism. We find that intermediate code distances reduce to a minimum of 7, while intermediate check weights reach a maximum of 16.

For completeness, we apply the inverse map to the balanced product code polynomials described in \cref{eqn: BP representative polynomials}. We get the bivariate bicycle codes defined by 
\begin{subequations}
\begin{align}
    A' &= 1+\alpha+\alpha^5,\\
    B' &= 1+ \alpha^{12}\beta+\alpha^{12}\beta^2.
\end{align}
\end{subequations}
\section{Conclusion}\label{sec:conclusion}
In this work, we have generalised Dehn twists from topological codes to qLDPC codes constructed from the hypergraph and balanced products of classical cyclic codes. By describing the toric code using cyclic codes polynomials and homological algebra, we have repurposed Dehn twists from topology into a more general tool for implementing entangling logical operations. The generalised Dehn twists can be implemented
with no additional ancillary qubits, equipping these codes with a set of low-overhead entangling logical operations. Because error correction can be done in-between steps, the generalised Dehn twists are also fault-tolerant if appropriate representative polynomials can be chosen, such that the check weights of intermediate codes are sufficiently low, and intermediate distances are sufficiently high. The extension to the general code family also means that there is room to control distance and logical qubit numbers, giving more manoeuvrability for experimental implementation.

To illustrate the generalised Dehn twists, we have focused on the hypergraph and balanced product codes that scale as $[[18q^2,8,2q]]$ and $[[18q,8,\leq 2q]]$ respectively. %

For the hypergraph product code, we have found that the set of Dehn twists is sufficient to generate the full logical Clifford group when supplemented with fold-transversal gates for codes with distance $d\in\{6,8,10,12\}$. We have also shown numerically that implementing the Dehn twists for some check polynomials that intermediate code distances do not reduce; and that maximum intermediate check weights are 9. 

For the balanced product code, we have shown that the set of Dehn twists apply to codes with odd $q$, due to an odd overlap with the twist columns. We have also shown that codes constructed in this way, taking the balanced product over a cyclic subgroup, have the topology of a twisted torus. Focusing on an example $[[90,8,10]]$ balanced product code, we found that, unlike the hypergraph product code, the set of Dehn twists and fold-transversal gates is not sufficient to generate the full logical Clifford group. Implementing Dehn twists numerically, we also found that intermediate code distances reduce to a minimum of 6.

We have also identified an isomorphism between this example code family and the $[[90,8,10]]$ bivariate bicycle code. This extends the generalised Dehn twists and the logical basis to the code, the latter of which was previously found in \cite{bravyi2023highthreshold} through numerical simulation. Moreover, rewriting the bivariate bicycle codes as a balanced product highlights a connection to the starting hypergraph product code. This might make the code amenable to measurement scheduling schemes and other relevant discussions, such as those in \cite{PhysRevLett.129.050504,manes2023distancepreservingstabilizermeasurementshypergraph}.

The balanced product cyclic codes also have interesting properties in that they are small codes upper-bounded by a linear distance. On its own, the upper bound is not significant. However, we have found some codes that have saturated this distance bound, and results also suggest that increasing the check weight returns codes with distances that more closely approach the bound. Our discussion on preserving code homology provides a recipe for similar small balanced product cyclic codes to be constructed, by factoring out an appropriate subgroup $\langle x^j \rangle, \deg p(x)< j$ for cyclic subgroup $\langle x^j\rangle $ and check polynomial $p(x)$. This may lead to other small codes that saturate the distance upper bound. 

We also point out some complementary work published recently in \cite{swaroop2024universaladaptersquantumldpc}. In their work, the toric code is used as an ancillary system in a universal toric code \textit{adapter} to implement logical $\operatorname{CNOT}$ gates on arbitrary qLDPC codes via Dehn twists. 
Our work generalising the Dehn twists is complementary in the sense that we increase the number of logical qubits supported on the handle, and the number of logical operations available via Dehn twists. If there is a way to incorporate generalised Dehn twists into the universal adapters, it might be possible to combine the schemes to implement more logical gates on arbitary qLDPC codes.

As a final comment, our work also raises some natural pathways for further investigation. We list them here as potential avenues for future work.
\begin{itemize}
    \item To what extent can increasing polynomial weights increase code distance for the balanced product cyclic codes?
\end{itemize}
Some examples of the $[[18q,8,\leq2q]]$ balanced product cyclic code presented in this work have maximum distances that are not achievable using weight 3 polynomials. However, we have also shown that using combinations of weight 3 and weight 5 polynomial representatives sometimes allow us to saturate the distance bound. In other cases it might improve the code distance without saturating the bound. A natural question is therefore to ask whether codes that saturate the bound for \textit{all} examples from this family exist, and whether they can be achieved through combinations of different polynomials with higher weights. If so, this would be a method to construct small families of codes with linear distance, making it promising for current architecture.

\begin{itemize}
    \item Can Dehn twists be extended to other bivariate bicycle codes?
\end{itemize}
There are two paths to extend Dehn twists in this way. We showed that the $[[90,8,10]]$ bivariate bicycle code is equivalent to the balanced product of cyclic codes. Given the similarity in construction, we can attempt to determine if other bivariate bicycle codes are also balanced product cyclic codes. If the answer is in the affirmative, then the Dehn twist scheme and discussion presented in this work apply immediately. Moreover, it will equip the codes with a logical basis that can be written analytically, eliminating the need for a numerical search.

It is also known that all bivariate bicycle codes can be written as balanced product codes \cite{eberhardt2024logicaloperatorsfoldtransversalgates}. They come from the double complex where every term is defined by $\mathbb{F}_2G\otimes_GG,G=C_l\otimes C_m$. Therefore another pathway to extend the Dehn twists is to determine their effect on codes from the balanced product of this tensor product of cyclic group algebras.
\begin{itemize}
    \item Are any other balanced product cyclic codes also biplanar codes?
\end{itemize}
The $[[90,8,10]]$ bivariate bicycle code was shown to be a biplanar code \cite{bravyi2023highthreshold}, needing only two layers of non-crossing circuitry to implement all qubit connections. Biplanarity makes the code amenable to experimental implementation. We showed in this work an equivalence between this bivariate bicycle code and a balanced product of cyclic codes. We ask if there are any other examples of this balanced product code family that are bilayer, again given the similarity in construction.
\begin{itemize}
    \item What are the scaling parameters of the Dehn twists and code families when using other check polynomials in the code construction?
\end{itemize}
The check polynomial $1+x+x^2$ is the simplest polynomial after that of the classical repetition code, and using this polynomial in our construction returns a quantum code with 8 qubits. We showed that 16 Dehn twists generate the $\operatorname{PSL}(8,2)$ group on the logical space.%
Using different check polynomials can increase the number of twists, and in general it is not known how this affects the generated group. 
\begin{itemize}
    \item Can distance reduction for balanced product cyclic codes be solved analytically?
\end{itemize}
Work done in this paper is largely achieved by defining cyclic codes and Dehn twists in the language of polynomials and homological algebra. In the proposed balanced product cyclic codes, we showed through considering homology that the resultant code preserves the number of logical qubits. We can also frame the problem of distance reduction analytically in this context, rather than rely on brute-force numerical searches. Then a natural question is to ask if there as an analytic solution. If so, it eliminates the need for a numerical search, and provides an explicit construction method for small families of quantum codes.
\begin{acknowledgments}
We would like to thank Harry Zhou and Quynh T.\ Nguyen for insightful discussions. RT acknowledges support from UK EPSRC (EP/SO23607/1). This work was done in part while the authors were visiting the Simons Institute for the Theory of Computing, supported by DOE QSA grant \#FP00010905. This work was carried out using the computational facilities of the Advanced Computing Research Centre, University of Bristol - \url{http://www.bris.ac.uk/acrc/}. 
\end{acknowledgments}
\newpage
\bibliography{apssamp}%

\providecommand{\noopsort}[1]{}\providecommand{\singleletter}[1]{#1}%
\begin{thebibliography}{30}%
\makeatletter
\providecommand \@ifxundefined [1]{%
 \@ifx{#1\undefined}
}%
\providecommand \@ifnum [1]{%
 \ifnum #1\expandafter \@firstoftwo
 \else \expandafter \@secondoftwo
 \fi
}%
\providecommand \@ifx [1]{%
 \ifx #1\expandafter \@firstoftwo
 \else \expandafter \@secondoftwo
 \fi
}%
\providecommand \natexlab [1]{#1}%
\providecommand \enquote  [1]{``#1''}%
\providecommand \bibnamefont  [1]{#1}%
\providecommand \bibfnamefont [1]{#1}%
\providecommand \citenamefont [1]{#1}%
\providecommand \href@noop [0]{\@secondoftwo}%
\providecommand \href [0]{\begingroup \@sanitize@url \@href}%
\providecommand \@href[1]{\@@startlink{#1}\@@href}%
\providecommand \@@href[1]{\endgroup#1\@@endlink}%
\providecommand \@sanitize@url [0]{\catcode `\\12\catcode `\$12\catcode `\&12\catcode `\#12\catcode `\^12\catcode `\_12\catcode `\%12\relax}%
\providecommand \@@startlink[1]{}%
\providecommand \@@endlink[0]{}%
\providecommand \url  [0]{\begingroup\@sanitize@url \@url }%
\providecommand \@url [1]{\endgroup\@href {#1}{\urlprefix }}%
\providecommand \urlprefix  [0]{URL }%
\providecommand \Eprint [0]{\href }%
\providecommand \doibase [0]{https://doi.org/}%
\providecommand \selectlanguage [0]{\@gobble}%
\providecommand \bibinfo  [0]{\@secondoftwo}%
\providecommand \bibfield  [0]{\@secondoftwo}%
\providecommand \translation [1]{[#1]}%
\providecommand \BibitemOpen [0]{}%
\providecommand \bibitemStop [0]{}%
\providecommand \bibitemNoStop [0]{.\EOS\space}%
\providecommand \EOS [0]{\spacefactor3000\relax}%
\providecommand \BibitemShut  [1]{\csname bibitem#1\endcsname}%
\let\auto@bib@innerbib\@empty
\bibitem [{\citenamefont {Bravyi}\ \emph {et~al.}(2023)\citenamefont {Bravyi}, \citenamefont {Cross}, \citenamefont {Gambetta}, \citenamefont {Maslov}, \citenamefont {Rall},\ and\ \citenamefont {Yoder}}]{bravyi2023highthreshold}%
  \BibitemOpen
  \bibfield  {author} {\bibinfo {author} {\bibfnamefont {S.}~\bibnamefont {Bravyi}}, \bibinfo {author} {\bibfnamefont {A.~W.}\ \bibnamefont {Cross}}, \bibinfo {author} {\bibfnamefont {J.~M.}\ \bibnamefont {Gambetta}}, \bibinfo {author} {\bibfnamefont {D.}~\bibnamefont {Maslov}}, \bibinfo {author} {\bibfnamefont {P.}~\bibnamefont {Rall}},\ and\ \bibinfo {author} {\bibfnamefont {T.~J.}\ \bibnamefont {Yoder}},\ }\href@noop {} {\bibinfo {title} {High-threshold and low-overhead fault-tolerant quantum memory}} (\bibinfo {year} {2023}),\ \Eprint {https://arxiv.org/abs/2308.07915} {arXiv:2308.07915 [quant-ph]} \BibitemShut {NoStop}%
\bibitem [{\citenamefont {Breuckmann}\ and\ \citenamefont {Eberhardt}(2021{\natexlab{a}})}]{Breuckmann_2021}%
  \BibitemOpen
  \bibfield  {author} {\bibinfo {author} {\bibfnamefont {N.~P.}\ \bibnamefont {Breuckmann}}\ and\ \bibinfo {author} {\bibfnamefont {J.~N.}\ \bibnamefont {Eberhardt}},\ }\bibfield  {title} {\bibinfo {title} {Quantum low-density parity-check codes},\ }\bibfield  {journal} {\bibinfo  {journal} {{PRX} Quantum}\ }\textbf {\bibinfo {volume} {2}},\ \href {https://doi.org/10.1103/prxquantum.2.040101} {10.1103/prxquantum.2.040101} (\bibinfo {year} {2021}{\natexlab{a}})\BibitemShut {NoStop}%
\bibitem [{\citenamefont {Acharya}\ \emph {et~al.}(2024)\citenamefont {Acharya}, \citenamefont {Aghababaie-Beni}, \citenamefont {Aleiner}, \citenamefont {Andersen}, \citenamefont {Ansmann}, \citenamefont {Arute}, \citenamefont {Arya}, \citenamefont {Asfaw}, \citenamefont {Astrakhantsev}, \citenamefont {Atalaya}, \citenamefont {Babbush}, \citenamefont {Bacon}, \citenamefont {Ballard}, \citenamefont {Bardin}, \citenamefont {Bausch} \emph {et~al.}}]{acharya2024quantumerrorcorrectionsurface}%
  \BibitemOpen
  \bibfield  {author} {\bibinfo {author} {\bibfnamefont {R.}~\bibnamefont {Acharya}}, \bibinfo {author} {\bibfnamefont {L.}~\bibnamefont {Aghababaie-Beni}}, \bibinfo {author} {\bibfnamefont {I.}~\bibnamefont {Aleiner}}, \bibinfo {author} {\bibfnamefont {T.~I.}\ \bibnamefont {Andersen}}, \bibinfo {author} {\bibfnamefont {M.}~\bibnamefont {Ansmann}}, \bibinfo {author} {\bibfnamefont {F.}~\bibnamefont {Arute}}, \bibinfo {author} {\bibfnamefont {K.}~\bibnamefont {Arya}}, \bibinfo {author} {\bibfnamefont {A.}~\bibnamefont {Asfaw}}, \bibinfo {author} {\bibfnamefont {N.}~\bibnamefont {Astrakhantsev}}, \bibinfo {author} {\bibfnamefont {J.}~\bibnamefont {Atalaya}}, \bibinfo {author} {\bibfnamefont {R.}~\bibnamefont {Babbush}}, \bibinfo {author} {\bibfnamefont {D.}~\bibnamefont {Bacon}}, \bibinfo {author} {\bibfnamefont {B.}~\bibnamefont {Ballard}}, \bibinfo {author} {\bibfnamefont {J.~C.}\ \bibnamefont {Bardin}}, \bibinfo {author} {\bibfnamefont {J.}~\bibnamefont {Bausch}}, \emph {et~al.},\ }\href
  {https://arxiv.org/abs/2408.13687} {\bibinfo {title} {Quantum error correction below the surface code threshold}} (\bibinfo {year} {2024}),\ \Eprint {https://arxiv.org/abs/2408.13687} {arXiv:2408.13687 [quant-ph]} \BibitemShut {NoStop}%
\bibitem [{\citenamefont {Bluvstein}\ \emph {et~al.}(2024)\citenamefont {Bluvstein}, \citenamefont {Evered}, \citenamefont {Geim}, \citenamefont {Li}, \citenamefont {Zhou}, \citenamefont {Manovitz}, \citenamefont {Ebadi}, \citenamefont {Cain}, \citenamefont {Kalinowski}, \citenamefont {Hangleiter}, \citenamefont {Bonilla~Ataides}, \citenamefont {Maskara}, \citenamefont {Cong}, \citenamefont {Gao}, \citenamefont {Sales~Rodriguez} \emph {et~al.}}]{Bluvstein2024}%
  \BibitemOpen
  \bibfield  {author} {\bibinfo {author} {\bibfnamefont {D.}~\bibnamefont {Bluvstein}}, \bibinfo {author} {\bibfnamefont {S.~J.}\ \bibnamefont {Evered}}, \bibinfo {author} {\bibfnamefont {A.~A.}\ \bibnamefont {Geim}}, \bibinfo {author} {\bibfnamefont {S.~H.}\ \bibnamefont {Li}}, \bibinfo {author} {\bibfnamefont {H.}~\bibnamefont {Zhou}}, \bibinfo {author} {\bibfnamefont {T.}~\bibnamefont {Manovitz}}, \bibinfo {author} {\bibfnamefont {S.}~\bibnamefont {Ebadi}}, \bibinfo {author} {\bibfnamefont {M.}~\bibnamefont {Cain}}, \bibinfo {author} {\bibfnamefont {M.}~\bibnamefont {Kalinowski}}, \bibinfo {author} {\bibfnamefont {D.}~\bibnamefont {Hangleiter}}, \bibinfo {author} {\bibfnamefont {J.~P.}\ \bibnamefont {Bonilla~Ataides}}, \bibinfo {author} {\bibfnamefont {N.}~\bibnamefont {Maskara}}, \bibinfo {author} {\bibfnamefont {I.}~\bibnamefont {Cong}}, \bibinfo {author} {\bibfnamefont {X.}~\bibnamefont {Gao}}, \bibinfo {author} {\bibfnamefont {P.}~\bibnamefont {Sales~Rodriguez}}, \emph {et~al.},\ }\bibfield  {title}
  {\bibinfo {title} {Logical quantum processor based on reconfigurable atom arrays},\ }\href {https://doi.org/10.1038/s41586-023-06927-3} {\bibfield  {journal} {\bibinfo  {journal} {Nature}\ }\textbf {\bibinfo {volume} {626}},\ \bibinfo {pages} {58} (\bibinfo {year} {2024})}\BibitemShut {NoStop}%
\bibitem [{\citenamefont {Xu}\ \emph {et~al.}(2024)\citenamefont {Xu}, \citenamefont {Zhou}, \citenamefont {Zheng}, \citenamefont {Bluvstein}, \citenamefont {Ataides}, \citenamefont {Lukin},\ and\ \citenamefont {Jiang}}]{xu2024fastparallelizablelogicalcomputation}%
  \BibitemOpen
  \bibfield  {author} {\bibinfo {author} {\bibfnamefont {Q.}~\bibnamefont {Xu}}, \bibinfo {author} {\bibfnamefont {H.}~\bibnamefont {Zhou}}, \bibinfo {author} {\bibfnamefont {G.}~\bibnamefont {Zheng}}, \bibinfo {author} {\bibfnamefont {D.}~\bibnamefont {Bluvstein}}, \bibinfo {author} {\bibfnamefont {J.~P.~B.}\ \bibnamefont {Ataides}}, \bibinfo {author} {\bibfnamefont {M.~D.}\ \bibnamefont {Lukin}},\ and\ \bibinfo {author} {\bibfnamefont {L.}~\bibnamefont {Jiang}},\ }\href {https://arxiv.org/abs/2407.18490} {\bibinfo {title} {Fast and parallelizable logical computation with homological product codes}} (\bibinfo {year} {2024}),\ \Eprint {https://arxiv.org/abs/2407.18490} {arXiv:2407.18490 [quant-ph]} \BibitemShut {NoStop}%
\bibitem [{\citenamefont {Ryan-Anderson}\ \emph {et~al.}(2022)\citenamefont {Ryan-Anderson}, \citenamefont {Brown}, \citenamefont {Allman}, \citenamefont {Arkin}, \citenamefont {Asa-Attuah}, \citenamefont {Baldwin}, \citenamefont {Berg}, \citenamefont {Bohnet}, \citenamefont {Braxton}, \citenamefont {Burdick}, \citenamefont {Campora}, \citenamefont {Chernoguzov}, \citenamefont {Esposito}, \citenamefont {Evans}, \citenamefont {Francois} \emph {et~al.}}]{ryananderson2022implementingfaulttolerantentanglinggates}%
  \BibitemOpen
  \bibfield  {author} {\bibinfo {author} {\bibfnamefont {C.}~\bibnamefont {Ryan-Anderson}}, \bibinfo {author} {\bibfnamefont {N.~C.}\ \bibnamefont {Brown}}, \bibinfo {author} {\bibfnamefont {M.~S.}\ \bibnamefont {Allman}}, \bibinfo {author} {\bibfnamefont {B.}~\bibnamefont {Arkin}}, \bibinfo {author} {\bibfnamefont {G.}~\bibnamefont {Asa-Attuah}}, \bibinfo {author} {\bibfnamefont {C.}~\bibnamefont {Baldwin}}, \bibinfo {author} {\bibfnamefont {J.}~\bibnamefont {Berg}}, \bibinfo {author} {\bibfnamefont {J.~G.}\ \bibnamefont {Bohnet}}, \bibinfo {author} {\bibfnamefont {S.}~\bibnamefont {Braxton}}, \bibinfo {author} {\bibfnamefont {N.}~\bibnamefont {Burdick}}, \bibinfo {author} {\bibfnamefont {J.~P.}\ \bibnamefont {Campora}}, \bibinfo {author} {\bibfnamefont {A.}~\bibnamefont {Chernoguzov}}, \bibinfo {author} {\bibfnamefont {J.}~\bibnamefont {Esposito}}, \bibinfo {author} {\bibfnamefont {B.}~\bibnamefont {Evans}}, \bibinfo {author} {\bibfnamefont {D.}~\bibnamefont {Francois}}, \emph {et~al.},\ }\href
  {https://arxiv.org/abs/2208.01863} {\bibinfo {title} {Implementing fault-tolerant entangling gates on the five-qubit code and the color code}} (\bibinfo {year} {2022}),\ \Eprint {https://arxiv.org/abs/2208.01863} {arXiv:2208.01863 [quant-ph]} \BibitemShut {NoStop}%
\bibitem [{\citenamefont {Hong}\ \emph {et~al.}(2024)\citenamefont {Hong}, \citenamefont {Durso-Sabina}, \citenamefont {Hayes},\ and\ \citenamefont {Lucas}}]{PhysRevLett.133.180601}%
  \BibitemOpen
  \bibfield  {author} {\bibinfo {author} {\bibfnamefont {Y.}~\bibnamefont {Hong}}, \bibinfo {author} {\bibfnamefont {E.}~\bibnamefont {Durso-Sabina}}, \bibinfo {author} {\bibfnamefont {D.}~\bibnamefont {Hayes}},\ and\ \bibinfo {author} {\bibfnamefont {A.}~\bibnamefont {Lucas}},\ }\bibfield  {title} {\bibinfo {title} {Entangling four logical qubits beyond break-even in a nonlocal code},\ }\href {https://doi.org/10.1103/PhysRevLett.133.180601} {\bibfield  {journal} {\bibinfo  {journal} {Phys. Rev. Lett.}\ }\textbf {\bibinfo {volume} {133}},\ \bibinfo {pages} {180601} (\bibinfo {year} {2024})}\BibitemShut {NoStop}%
\bibitem [{\citenamefont {Jochym-O’Connor}(2019)}]{Jochym_O_Connor_2019}%
  \BibitemOpen
  \bibfield  {author} {\bibinfo {author} {\bibfnamefont {T.}~\bibnamefont {Jochym-O’Connor}},\ }\bibfield  {title} {\bibinfo {title} {Fault-tolerant gates via homological product codes},\ }\href {https://doi.org/10.22331/q-2019-02-04-120} {\bibfield  {journal} {\bibinfo  {journal} {Quantum}\ }\textbf {\bibinfo {volume} {3}},\ \bibinfo {pages} {120} (\bibinfo {year} {2019})}\BibitemShut {NoStop}%
\bibitem [{\citenamefont {Krishna}\ and\ \citenamefont {Poulin}(2021)}]{PhysRevX.11.011023}%
  \BibitemOpen
  \bibfield  {author} {\bibinfo {author} {\bibfnamefont {A.}~\bibnamefont {Krishna}}\ and\ \bibinfo {author} {\bibfnamefont {D.}~\bibnamefont {Poulin}},\ }\bibfield  {title} {\bibinfo {title} {Fault-tolerant gates on hypergraph product codes},\ }\href {https://doi.org/10.1103/PhysRevX.11.011023} {\bibfield  {journal} {\bibinfo  {journal} {Phys. Rev. X}\ }\textbf {\bibinfo {volume} {11}},\ \bibinfo {pages} {011023} (\bibinfo {year} {2021})}\BibitemShut {NoStop}%
\bibitem [{\citenamefont {Cohen}\ \emph {et~al.}(2022)\citenamefont {Cohen}, \citenamefont {Kim}, \citenamefont {Bartlett},\ and\ \citenamefont {Brown}}]{Cohen_2022}%
  \BibitemOpen
  \bibfield  {author} {\bibinfo {author} {\bibfnamefont {L.~Z.}\ \bibnamefont {Cohen}}, \bibinfo {author} {\bibfnamefont {I.~H.}\ \bibnamefont {Kim}}, \bibinfo {author} {\bibfnamefont {S.~D.}\ \bibnamefont {Bartlett}},\ and\ \bibinfo {author} {\bibfnamefont {B.~J.}\ \bibnamefont {Brown}},\ }\bibfield  {title} {\bibinfo {title} {Low-overhead fault-tolerant quantum computing using long-range connectivity},\ }\bibfield  {journal} {\bibinfo  {journal} {Science Advances}\ }\textbf {\bibinfo {volume} {8}},\ \href {https://doi.org/10.1126/sciadv.abn1717} {10.1126/sciadv.abn1717} (\bibinfo {year} {2022})\BibitemShut {NoStop}%
\bibitem [{\citenamefont {Breuckmann}\ and\ \citenamefont {Burton}(2022)}]{breuckmann2022foldtransversal}%
  \BibitemOpen
  \bibfield  {author} {\bibinfo {author} {\bibfnamefont {N.~P.}\ \bibnamefont {Breuckmann}}\ and\ \bibinfo {author} {\bibfnamefont {S.}~\bibnamefont {Burton}},\ }\href@noop {} {\bibinfo {title} {Fold-transversal clifford gates for quantum codes}} (\bibinfo {year} {2022}),\ \Eprint {https://arxiv.org/abs/2202.06647} {arXiv:2202.06647 [quant-ph]} \BibitemShut {NoStop}%
\bibitem [{\citenamefont {Quintavalle}\ \emph {et~al.}(2023)\citenamefont {Quintavalle}, \citenamefont {Webster},\ and\ \citenamefont {Vasmer}}]{Quintavalle2023partitioningqubits}%
  \BibitemOpen
  \bibfield  {author} {\bibinfo {author} {\bibfnamefont {A.~O.}\ \bibnamefont {Quintavalle}}, \bibinfo {author} {\bibfnamefont {P.}~\bibnamefont {Webster}},\ and\ \bibinfo {author} {\bibfnamefont {M.}~\bibnamefont {Vasmer}},\ }\bibfield  {title} {\bibinfo {title} {Partitioning qubits in hypergraph product codes to implement logical gates},\ }\href {https://doi.org/10.22331/q-2023-10-24-1153} {\bibfield  {journal} {\bibinfo  {journal} {{Quantum}}\ }\textbf {\bibinfo {volume} {7}},\ \bibinfo {pages} {1153} (\bibinfo {year} {2023})}\BibitemShut {NoStop}%
\bibitem [{\citenamefont {Cross}\ \emph {et~al.}(2024)\citenamefont {Cross}, \citenamefont {He}, \citenamefont {Rall},\ and\ \citenamefont {Yoder}}]{cross2024improvedqldpcsurgerylogical}%
  \BibitemOpen
  \bibfield  {author} {\bibinfo {author} {\bibfnamefont {A.}~\bibnamefont {Cross}}, \bibinfo {author} {\bibfnamefont {Z.}~\bibnamefont {He}}, \bibinfo {author} {\bibfnamefont {P.}~\bibnamefont {Rall}},\ and\ \bibinfo {author} {\bibfnamefont {T.}~\bibnamefont {Yoder}},\ }\href {https://arxiv.org/abs/2407.18393} {\bibinfo {title} {Improved qldpc surgery: Logical measurements and bridging codes}} (\bibinfo {year} {2024}),\ \Eprint {https://arxiv.org/abs/2407.18393} {arXiv:2407.18393 [quant-ph]} \BibitemShut {NoStop}%
\bibitem [{\citenamefont {Kitaev}(2003)}]{Kitaev_2003}%
  \BibitemOpen
  \bibfield  {author} {\bibinfo {author} {\bibfnamefont {A.}~\bibnamefont {Kitaev}},\ }\bibfield  {title} {\bibinfo {title} {Fault-tolerant quantum computation by anyons},\ }\href {https://doi.org/10.1016/s0003-4916(02)00018-0} {\bibfield  {journal} {\bibinfo  {journal} {Annals of Physics}\ }\textbf {\bibinfo {volume} {303}},\ \bibinfo {pages} {2–30} (\bibinfo {year} {2003})}\BibitemShut {NoStop}%
\bibitem [{\citenamefont {Koenig}\ \emph {et~al.}(2010)\citenamefont {Koenig}, \citenamefont {Kuperberg},\ and\ \citenamefont {Reichardt}}]{Koenig_2010}%
  \BibitemOpen
  \bibfield  {author} {\bibinfo {author} {\bibfnamefont {R.}~\bibnamefont {Koenig}}, \bibinfo {author} {\bibfnamefont {G.}~\bibnamefont {Kuperberg}},\ and\ \bibinfo {author} {\bibfnamefont {B.~W.}\ \bibnamefont {Reichardt}},\ }\bibfield  {title} {\bibinfo {title} {Quantum computation with turaev–viro codes},\ }\href {https://doi.org/10.1016/j.aop.2010.08.001} {\bibfield  {journal} {\bibinfo  {journal} {Annals of Physics}\ }\textbf {\bibinfo {volume} {325}},\ \bibinfo {pages} {2707–2749} (\bibinfo {year} {2010})}\BibitemShut {NoStop}%
\bibitem [{\citenamefont {Breuckmann}\ \emph {et~al.}(2017)\citenamefont {Breuckmann}, \citenamefont {Vuillot}, \citenamefont {Campbell}, \citenamefont {Krishna},\ and\ \citenamefont {Terhal}}]{Breuckmann_2017}%
  \BibitemOpen
  \bibfield  {author} {\bibinfo {author} {\bibfnamefont {N.~P.}\ \bibnamefont {Breuckmann}}, \bibinfo {author} {\bibfnamefont {C.}~\bibnamefont {Vuillot}}, \bibinfo {author} {\bibfnamefont {E.}~\bibnamefont {Campbell}}, \bibinfo {author} {\bibfnamefont {A.}~\bibnamefont {Krishna}},\ and\ \bibinfo {author} {\bibfnamefont {B.~M.}\ \bibnamefont {Terhal}},\ }\bibfield  {title} {\bibinfo {title} {Hyperbolic and semi-hyperbolic surface codes for quantum storage},\ }\href {https://doi.org/10.1088/2058-9565/aa7d3b} {\bibfield  {journal} {\bibinfo  {journal} {Quantum Science and Technology}\ }\textbf {\bibinfo {volume} {2}},\ \bibinfo {pages} {035007} (\bibinfo {year} {2017})}\BibitemShut {NoStop}%
\bibitem [{\citenamefont {Breuckmann}(2018)}]{breuckmann2018phdthesishomologicalquantum}%
  \BibitemOpen
  \bibfield  {author} {\bibinfo {author} {\bibfnamefont {N.~P.}\ \bibnamefont {Breuckmann}},\ }\href {https://arxiv.org/abs/1802.01520} {\bibinfo {title} {Phd thesis: Homological quantum codes beyond the toric code}} (\bibinfo {year} {2018}),\ \Eprint {https://arxiv.org/abs/1802.01520} {arXiv:1802.01520 [quant-ph]} \BibitemShut {NoStop}%
\bibitem [{\citenamefont {Lavasani}\ \emph {et~al.}(2019)\citenamefont {Lavasani}, \citenamefont {Zhu},\ and\ \citenamefont {Barkeshli}}]{Lavasani_2019}%
  \BibitemOpen
  \bibfield  {author} {\bibinfo {author} {\bibfnamefont {A.}~\bibnamefont {Lavasani}}, \bibinfo {author} {\bibfnamefont {G.}~\bibnamefont {Zhu}},\ and\ \bibinfo {author} {\bibfnamefont {M.}~\bibnamefont {Barkeshli}},\ }\bibfield  {title} {\bibinfo {title} {Universal logical gates with constant overhead: instantaneous dehn twists for hyperbolic quantum codes},\ }\href {https://doi.org/10.22331/q-2019-08-26-180} {\bibfield  {journal} {\bibinfo  {journal} {Quantum}\ }\textbf {\bibinfo {volume} {3}},\ \bibinfo {pages} {180} (\bibinfo {year} {2019})}\BibitemShut {NoStop}%
\bibitem [{\citenamefont {Farb}\ and\ \citenamefont {Margalit}(2017)}]{Farb_Margalit_2017}%
  \BibitemOpen
  \bibfield  {author} {\bibinfo {author} {\bibfnamefont {B.}~\bibnamefont {Farb}}\ and\ \bibinfo {author} {\bibfnamefont {D.}~\bibnamefont {Margalit}},\ }\href@noop {} {\emph {\bibinfo {title} {A Primer on mapping class groups}}}\ (\bibinfo  {publisher} {Princeton University Press},\ \bibinfo {year} {2017})\BibitemShut {NoStop}%
\bibitem [{\citenamefont {Tillich}\ and\ \citenamefont {Zemor}(2014)}]{Tillich_2014}%
  \BibitemOpen
  \bibfield  {author} {\bibinfo {author} {\bibfnamefont {J.-P.}\ \bibnamefont {Tillich}}\ and\ \bibinfo {author} {\bibfnamefont {G.}~\bibnamefont {Zemor}},\ }\bibfield  {title} {\bibinfo {title} {Quantum ldpc codes with positive rate and minimum distance proportional to the square root of the blocklength},\ }\href {https://doi.org/10.1109/tit.2013.2292061} {\bibfield  {journal} {\bibinfo  {journal} {IEEE Transactions on Information Theory}\ }\textbf {\bibinfo {volume} {60}},\ \bibinfo {pages} {1193–1202} (\bibinfo {year} {2014})}\BibitemShut {NoStop}%
\bibitem [{\citenamefont {Breuckmann}\ and\ \citenamefont {Eberhardt}(2021{\natexlab{b}})}]{Breuckmann_2021_2}%
  \BibitemOpen
  \bibfield  {author} {\bibinfo {author} {\bibfnamefont {N.~P.}\ \bibnamefont {Breuckmann}}\ and\ \bibinfo {author} {\bibfnamefont {J.~N.}\ \bibnamefont {Eberhardt}},\ }\bibfield  {title} {\bibinfo {title} {Balanced product quantum codes},\ }\href {https://doi.org/10.1109/tit.2021.3097347} {\bibfield  {journal} {\bibinfo  {journal} {{IEEE} Transactions on Information Theory}\ }\textbf {\bibinfo {volume} {67}},\ \bibinfo {pages} {6653} (\bibinfo {year} {2021}{\natexlab{b}})}\BibitemShut {NoStop}%
\bibitem [{\citenamefont {Gottesman}(2006)}]{GOTTESMAN2006196}%
  \BibitemOpen
  \bibfield  {author} {\bibinfo {author} {\bibfnamefont {D.}~\bibnamefont {Gottesman}},\ }\bibfield  {title} {\bibinfo {title} {Quantum error correction and fault tolerance},\ }in\ \href {https://doi.org/https://doi.org/10.1016/B0-12-512666-2/00273-X} {\emph {\bibinfo {booktitle} {Encyclopedia of Mathematical Physics}}},\ \bibinfo {editor} {edited by\ \bibinfo {editor} {\bibfnamefont {J.-P.}\ \bibnamefont {Françoise}}, \bibinfo {editor} {\bibfnamefont {G.~L.}\ \bibnamefont {Naber}},\ and\ \bibinfo {editor} {\bibfnamefont {T.~S.}\ \bibnamefont {Tsun}}}\ (\bibinfo  {publisher} {Academic Press},\ \bibinfo {address} {Oxford},\ \bibinfo {year} {2006})\ pp.\ \bibinfo {pages} {196--201}\BibitemShut {NoStop}%
\bibitem [{\citenamefont {Massey}(2019)}]{massey2019basic}%
  \BibitemOpen
  \bibfield  {author} {\bibinfo {author} {\bibfnamefont {W.~S.}\ \bibnamefont {Massey}},\ }\href@noop {} {\emph {\bibinfo {title} {A basic course in algebraic topology}}},\ Vol.\ \bibinfo {volume} {127}\ (\bibinfo  {publisher} {Springer},\ \bibinfo {year} {2019})\BibitemShut {NoStop}%
\bibitem [{\citenamefont {MacWilliams}\ and\ \citenamefont {Sloane}(1978)}]{bok:MW}%
  \BibitemOpen
  \bibfield  {author} {\bibinfo {author} {\bibfnamefont {F.}~\bibnamefont {MacWilliams}}\ and\ \bibinfo {author} {\bibfnamefont {N.}~\bibnamefont {Sloane}},\ }\href@noop {} {\emph {\bibinfo {title} {The Theory of Error-Correcting Codes}}},\ \bibinfo {edition} {2nd}\ ed.\ (\bibinfo  {publisher} {North-holland Publishing Company},\ \bibinfo {year} {1978})\BibitemShut {NoStop}%
\bibitem [{\citenamefont {{Gurobi Optimization, LLC}}(2024)}]{gurobi}%
  \BibitemOpen
  \bibfield  {author} {\bibinfo {author} {\bibnamefont {{Gurobi Optimization, LLC}}},\ }\href {https://www.gurobi.com} {\bibinfo {title} {{Gurobi Optimizer Reference Manual}}} (\bibinfo {year} {2024})\BibitemShut {NoStop}%
\bibitem [{\citenamefont {Tiew}(2024)}]{ryan_tiew_2024_14041286}%
  \BibitemOpen
  \bibfield  {author} {\bibinfo {author} {\bibfnamefont {R.}~\bibnamefont {Tiew}},\ }\href {https://doi.org/10.5281/zenodo.14041286} {\bibinfo {title} {h1010134/balanced-product-cyclic-codes: v1}} (\bibinfo {year} {2024})\BibitemShut {NoStop}%
\bibitem [{\citenamefont {Tremblay}\ \emph {et~al.}(2022)\citenamefont {Tremblay}, \citenamefont {Delfosse},\ and\ \citenamefont {Beverland}}]{PhysRevLett.129.050504}%
  \BibitemOpen
  \bibfield  {author} {\bibinfo {author} {\bibfnamefont {M.~A.}\ \bibnamefont {Tremblay}}, \bibinfo {author} {\bibfnamefont {N.}~\bibnamefont {Delfosse}},\ and\ \bibinfo {author} {\bibfnamefont {M.~E.}\ \bibnamefont {Beverland}},\ }\bibfield  {title} {\bibinfo {title} {Constant-overhead quantum error correction with thin planar connectivity},\ }\href {https://doi.org/10.1103/PhysRevLett.129.050504} {\bibfield  {journal} {\bibinfo  {journal} {Phys. Rev. Lett.}\ }\textbf {\bibinfo {volume} {129}},\ \bibinfo {pages} {050504} (\bibinfo {year} {2022})}\BibitemShut {NoStop}%
\bibitem [{\citenamefont {Manes}\ and\ \citenamefont {Claes}(2023)}]{manes2023distancepreservingstabilizermeasurementshypergraph}%
  \BibitemOpen
  \bibfield  {author} {\bibinfo {author} {\bibfnamefont {A.~G.}\ \bibnamefont {Manes}}\ and\ \bibinfo {author} {\bibfnamefont {J.}~\bibnamefont {Claes}},\ }\href {https://arxiv.org/abs/2308.15520} {\bibinfo {title} {Distance-preserving stabilizer measurements in hypergraph product codes}} (\bibinfo {year} {2023}),\ \Eprint {https://arxiv.org/abs/2308.15520} {arXiv:2308.15520 [quant-ph]} \BibitemShut {NoStop}%
\bibitem [{\citenamefont {Swaroop}\ \emph {et~al.}(2024)\citenamefont {Swaroop}, \citenamefont {Jochym-O'Connor},\ and\ \citenamefont {Yoder}}]{swaroop2024universaladaptersquantumldpc}%
  \BibitemOpen
  \bibfield  {author} {\bibinfo {author} {\bibfnamefont {E.}~\bibnamefont {Swaroop}}, \bibinfo {author} {\bibfnamefont {T.}~\bibnamefont {Jochym-O'Connor}},\ and\ \bibinfo {author} {\bibfnamefont {T.~J.}\ \bibnamefont {Yoder}},\ }\href {https://arxiv.org/abs/2410.03628} {\bibinfo {title} {Universal adapters between quantum ldpc codes}} (\bibinfo {year} {2024}),\ \Eprint {https://arxiv.org/abs/2410.03628} {arXiv:2410.03628 [quant-ph]} \BibitemShut {NoStop}%
\bibitem [{\citenamefont {Eberhardt}\ and\ \citenamefont {Steffan}(2024)}]{eberhardt2024logicaloperatorsfoldtransversalgates}%
  \BibitemOpen
  \bibfield  {author} {\bibinfo {author} {\bibfnamefont {J.~N.}\ \bibnamefont {Eberhardt}}\ and\ \bibinfo {author} {\bibfnamefont {V.}~\bibnamefont {Steffan}},\ }\href {https://arxiv.org/abs/2407.03973} {\bibinfo {title} {Logical operators and fold-transversal gates of bivariate bicycle codes}} (\bibinfo {year} {2024}),\ \Eprint {https://arxiv.org/abs/2407.03973} {arXiv:2407.03973 [quant-ph]} \BibitemShut {NoStop}%
\end{thebibliography}%
\end{document}